\documentstyle[aps,prb,eqsecnum]{revtex}

\newcommand{\diag}{{\rm diag\,}}
\newcommand{\tr}{{\rm tr\,}}
\newcommand{\Tr}{{\rm Tr\,}}
\newcommand{\Det}{{\rm Det\,}}

\begin{document}

\title{Recursive Construction for a Class of Radial 
       Functions I --- Ordinary Space}

\author{Thomas Guhr\thanks{e-mail: guhr@daniel.mpi-hd.mpg.de}
        and
        Heiner Kohler\thanks{e-mail: kohler@daniel.mpi-hd.mpg.de}\\
        Max Planck Institut f\"ur Kernphysik,
        Postfach 103980,
        69029 Heidelberg,
        Germany}

\maketitle

\begin{abstract}
  A class of spherical functions is studied which can be viewed as the
  matrix generalization of Bessel functions. We derive a recursive
  structure for these functions. We show that they are only special
  cases of more general radial functions which also have a, properly
  generalized, recursive structure.  Some explicit results are worked
  out.
\end{abstract}

\section{Introduction}
\label{sec0}

In 1957, Harish--Chandra~\cite{HC1} derived a famous formula for a
certain class of group integrals. Let $\cal{G}$ be a compact
semi--simple Lie group and let $a$ and $b$ be elements of its
Cartan subalgebra $\cal{H}$, then
\begin{equation} 
\int_{U\in\cal{G}} d\mu(U) \exp\left(\tr U^{-1}a Ub\right) 
        \ = \ \frac{1}{|\cal{W}|} \, 
              \sum_{w\in\cal{W}}
              \frac{\exp\left(\tr w(a)b\right)}
                   {\Pi(a)\Pi(w(b))} \ . 
\label{hcf}
\end{equation}
Here, $d\mu(U)$ is the invariant measure, $\Pi(a)$ is the product of
all positive roots of $\cal{H}$ and $\cal{W}$ is the Weyl
reflection group of $\cal{G}$ with $|\cal{W}|$ elements $w$.

This result depends crucially on the condition that $a$ and $b$ are in
the Cartan subalgebra $\cal{H}$. In other words, $U^{-1}a Ub$
has to be in $\cal{G}$. If one replaces $a$ and $b$ in the
integral on the left hand side with more general matrices $x$ and $k$
which are not in $\cal{H}$, formula~(\ref{hcf}) is {\it not
  valid anymore}.  The {\it spherical functions} introduced by
Gelfand~\cite{GEL1,HEL} form an important class of such integrals which
are, in general, not covered by Harish--Chandra's result~(\ref{hcf}).
In another work, Harish--Chandra~\cite{HC} studies in great detail the
harmonic analysis involving these spherical functions. In a more
physics oriented contribution, Olshanetsky and Perelomov~\cite{OP}
discussed them in the framework of quantum integrable systems.

Here, we wish to address spherical functions of the following kind: we
take $x$ and $k$ as diagonal matrices containing the eigenvalues of a
Hamiltonian in a matrix representation. The Hamiltonian is
diagonalized by the integration matrix $U$. In particular, we assume
that the Hamiltonian $U^{-1}xU$ or, equivalently, $UkU^{-1}$ is
real--symmetric, Hermitean or Hermitean self--dual. Thus,
$\cal{G}$ is the orthogonal, the unitary or the
unitary--symplectic group. We will refer to these spherical functions
as {\it matrix Bessel functions}.  We notice that the unitary case is
special: since it so happens that the eigenvalues $x$ and $k$ do lie
in the Cartan subalgebra $\cal{H}$, the result~(\ref{hcf})
applies and coincides with the Itzykson--Zuber formula~\cite{IZ}.  In
the orthogonal and the unitary--symplectic cases, however,
formula~(\ref{hcf}) is not valid.

We choose the term matrix Bessel function for the spherical functions
to be discussed here, because they can be viewed as a natural
extension of the ordinary vector Bessel functions. However, due to the
rich features of the spherical functions, other extensions relating to
ordinary Bessel functions are equally natural.  Related functions have
been discussed and terms similar to matrix Bessel functions have
already been used by Hertz~\cite{HER}, by Gross and
Kunze~\cite{GK1,GK2}, by Holman~\cite{HOL} and by Okounkov and
Olshanski~\cite{OO}.  Kontsevich~\cite{KON} introduced the matrix Airy
functions.

Duistermaat and Heckman~\cite{DH} developed a stationary phase approach
involving localization for a class of spherical functions,
see also the treatise by Szabo~\cite{SZA}.

Remarkably, our matrix Bessel functions are only special cases of more
general objects which we call {\it radial functions}.  Moreover, there
is an important connection to the Calogero--Sutherland models which we
will discuss separately, see below.

The matrix Bessel functions are of considerable interest for
applications in physics. They appear in Random Matrix
Theory~\cite{Mehta,Haake,GMGW} which models spectral fluctuations of
complex systems, such as quantum chaotic ones. In particular, they are
the kernels of Dyson's Brownian motion~\cite{DYS1,DYS2} describing
crossover transitions between different symmetry or invariance
classes. Unfortunately, only the case of broken time--reversal
invariance can be treated explicitly with the help of the
Itzykson--Zuber formula. In the physically important cases of
conserved time--reversal invariance, the kernels are not known
analytically, as argued above. Muirhead~\cite{MUI} discusses spherical
functions in the framework of multivariate statistical theory.  In his
book, an expansion in terms of Jack polynomials for the orthogonal
case can be found. Such an expansion for arbitrary Dyson index was
recently worked out by Okounkov and Olshanski~\cite{OO}.

The goal of the present paper is to explore the structure of the
radial functions which contain the matrix Bessel functions as special
cases. In particular, we show how explicit results can be obtained.
The paper is organized as follows. In Sec.~\ref{sec1}, we briefly
review some properties of the vector Bessel functions. In doing so we
wish to help the reader in developing an intuition for the matrix
Bessel functions which we introduce in Sec.~\ref{sec2}. In
Sec.~\ref{sec4}, we state and derive a fundamental recursive structure
for matrix Bessel functions. We show in Sec.~\ref{sec7} that this
recursion is an iterative solution of general radial functions which
contain group integrals defining the matrix Bessel functions as
special case.  Secs.~\ref{sec4} and~\ref{sec7} are our main results.
In Sec.~\ref{sec8} we illustrate how the recursion can lead to closed
and explicit formulae. Because of its special importance, we discuss
the connection to Calogero--Sutherland models separately in
Sec.~\ref{sec9}. In Sec.~\ref{sec10}, we summarize and conclude.
Various aspects and calculations are collected in the appendix.

\section{Vector Bessel Functions Revisited}
\label{sec1}

Before turning to the matrix case, we compile, for the convenience of
the reader, some well known results for the vector case.

In a real, $d$ dimensional space with $d=2,3,4,\ldots$, we consider a
position vector $\vec{r}=(x_1,\ldots,x_d)$ and a wave vector
$\vec{k}=(k_1,\ldots,k_d)$. The plane wave
$\exp\left(i\vec{k}\cdot\vec{r}\right)$ satisfies the wave equation
\begin{equation}
\Delta \exp\left(i\vec{k}\cdot\vec{r}\right) \ = \ - \vec{k}^2 \,
              \exp\left(i\vec{k}\cdot\vec{r}\right) 
\label{eq1.1}
\end{equation}
where we define the Laplacean as in the physics literature, 
\begin{equation}
\Delta \ = \ \frac{\partial^2}{\partial\vec{r}^2}
       \ = \ \sum_{i=1}^d \frac{\partial^2}{\partial x_i^2} \ .
\label{eq1.2}
\end{equation}
The zeroth order Bessel function in this space is 
the angular average of the plane wave,
\begin{equation}
\chi^{(d)}(kr) \ = \ \int d\Omega \, 
                     \exp\left(i\vec{k}\cdot\vec{r}\right) \ , 
\label{eq1.3}
\end{equation}
over the solid angle $\Omega$, defining the orientation of either
$\vec{r}$ or $\vec{k}$.  In our context, it is advantageous to take
$\Omega$ as the solid angle of $\vec{k}$. Obviously, only the relative
angle between $\vec{r}$ and $\vec{k}$ matters and $\chi^{(d)}(kr)$ can
only depend on the product of the lengths $r=|\vec{r}|$ and
$k=|\vec{k}|$ of the two vectors. We normalize the measure $d\Omega$
with the volume $2\pi^{d/2}/\Gamma(d/2)$ of the unit sphere, i.e.~we
have
\begin{equation}
\int d\Omega \ = \ 1 \ . 
\label{eq1.4}
\end{equation}
Thus, by construction, we also have
\begin{equation}
\chi^{(d)}(0) \ = \ 1 \ .
\label{eq1.5}
\end{equation}
It is convenient to view $\vec{r}$ as the azimuthal direction of the
coordinate system in which we measure $\Omega$. Thus, in these
spherical coordinates, one finds $\vec{k}\cdot\vec{r}=kr\cos\vartheta$
where $\vartheta$ is the azimuthal angle. The measure $d\Omega$
contains $\sin^{d-2}\vartheta$ and one has
\begin{eqnarray}
\chi^{(d)}(kr) &=& \frac{\Gamma(d/2)}{\sqrt{\pi}\Gamma((d-1)/2)} \,
                     \int_0^\pi \exp\left(ikr\cos\vartheta\right) 
                     \sin^{d-2}\vartheta d\vartheta
                                \nonumber\\
               &=& 2^{(d-2)/2} \Gamma(d/2) \, 
                     \frac{J_{(d-2)/2}(kr)}{(kr)^{(d-2)/2}}
\label{eq1.6}
\end{eqnarray}
where $J_\nu(z)$ is the standard Bessel function~\cite{AS} of order
$\nu$. The functions~(\ref{eq1.6}) are often referred to as zonal
functions.  

There is a remarkable difference for the functions $\chi^{(d)}(kr)$ if
one compares even and odd dimensions. For example, one has in $d=2$
dimensions $\chi^{(2)}(kr)= J_0(kr)$ and in $d=3$ dimensions
$\chi^{(3)}(kr)=(\pi/2)^{1/2} J_{1/2}(kr)/(kr)^{1/2} = j_0(kr)$ with
the spherical Bessel function $j_0(z)$ of zeroth order~\cite{AS}.  In
$d=2$ dimensions, $J_0(z)$ is a complicated infinite series in the
argument $z$, in $d=3$ dimensions, however, $j_0(z)$ is the simple
ratio $j_0(z)=\sin z/z$. One easily sees how this generalizes. Upon
introducing $\xi=\cos\vartheta$ as integration variable in
Eq.~(\ref{eq1.6}), one finds the representation
\begin{equation}
\chi^{(d)}(kr) \ = \ \frac{\Gamma(d/2)}{\sqrt{\pi}\Gamma((d-1)/2)} \,
                     \int_{-1}^{+1} \exp\left(ikr\xi\right) 
                     \left(1-\xi^2\right)^{(d-3)/2} d\xi \ .
\label{eq1.7}
\end{equation}
In dimensions $d\ge 3$, this can be cast into the form
\begin{equation}
\chi^{(d)}(kr) \ = \ \frac{2\Gamma(d/2)}{\sqrt{\pi}\Gamma((d-1)/2)} \,
                   \sum_{\mu=0}^\infty \left( \begin{array}{c}
                                              (d-3)/2 \\
                                              \mu
                                              \end{array} \right)
                   \frac{\partial^{2\mu}}{\partial(kr)^{2\mu}}
                   \frac{\sin kr}{kr} \ .
\label{eq1.8}
\end{equation}
For even $d$, the exponent $(d-3)/2$ is a fraction
$-1/2,+1/2,+3/2,\ldots$, and the function
$\left(1-\xi^2\right)^{(d-3)/2}$ in the integrand in Eq.~(\ref{eq1.7})
is an {\it infinite} power series. This yields, for $d=4,6,8,\ldots$,
the complicated power series~(\ref{eq1.8}) involving an infinite
number of inverse powers of $kr$. However, if $d$ is odd, the exponent
$(d-3)/2$ is an integer $0,1,2,\ldots$, and the function
$\left(1-\xi^2\right)^{(d-3)/2}$ is a {\it finite} polynomial of order
$(d-3)/2$ in $\xi^2$. Thus, $\chi^{(d)}(kr)$ acquires a comparatively
simple structure, because it only contains a finite number of inverse
powers of $kr$.  Formally, this means that for odd $d$ all binomial
coefficients for $\mu>(d-3)/2$ are zero.

The differential equation for the functions $\chi^{(d)}(kr)$ is easily
obtained by averaging Eq.~(\ref{eq1.1}) over the solid angle $\Omega$
of $\vec{k}$, i.e.~by integrating both sides,
\begin{equation}
\Delta \int d\Omega\exp\left(i\vec{k}\cdot\vec{r}\right) \ = \
         - \vec{k}^2 \int d\Omega
              \exp\left(i\vec{k}\cdot\vec{r}\right) \ .
\label{eq1.9}
\end{equation}
We notice that the Laplacean $\Delta$ commutes with the integral,
because the former is in the space of the position vector, the latter
in the space of the wave vector. Moreover, the integral trivially
commutes with $\vec{k}^2=k^2$. Hence, one arrives at
\begin{equation}
\Delta_r \chi^{(d)}(kr) \ = \ - k^2 \, \chi^{(d)}(kr) \ .
\label{eq1.10}
\end{equation}
Since $\chi^{(d)}(kr)$ depends exclusively on radial variables, we 
replaced the full La\-pla\-ce\-an $\Delta$ with its radial part
\begin{equation}
\Delta_r \ = \ \frac{1}{r^{d-1}}\frac{\partial}{\partial r}
               r^{d-1}\frac{\partial}{\partial r}
         \ = \ \frac{\partial^2}{\partial r^2} +
               \frac{d-1}{r}\frac{\partial}{\partial r} \ .
\label{eq1.11}
\end{equation}
In general, there are two fundamental solutions $\chi_+^{(d)}(kr)$ and
$\chi_-^{(d)}(kr)$ of the differential equation~(\ref{eq1.10}) which
behave as $\exp\left(\pm ikr\right)/\left(kr\right)^{(d-1)/2}$ for
large arguments $kr$. Thus, to obtain the full solutions, one can make
the Hankel ansatz
\begin{equation}
\chi_\pm^{(d)}(kr) \ = \ \frac{\exp\left(\pm ikr\right)}
                              {\left(kr\right)^{(d-1)/2}} \,
                              w_\pm^{(d)}(kr) \ .
\label{eq1.12}
\end{equation}
Here, $w_\pm^{(d)}(kr)$ is a function with the property
$w_\pm^{(d)}(kr)\to 1$ for $kr\to\infty$. The differential equation
follows easily from Eq.~(\ref{eq1.11}) and is given by
\begin{equation}
\left(\frac{\partial^2}{\partial r^2} 
     \pm i2k\frac{\partial}{\partial r}
     -\frac{d-1}{2}\left(\frac{d-1}{2}-1\right)\frac{1}{r^2}
                 \right)w_\pm^{(d)}(kr) \ = \ 0 \ .
\label{eq1.13}
\end{equation}
For $d\ge 3$, one uses the ansatz as an asymptotic power series 
\begin{equation}
w_\pm^{(d)}(kr) \ = \ \sum_{\mu=0}^\infty\frac{a_\mu}{(\pm kr)^\mu}
\label{eq1.14}
\end{equation}
which yields a recursion for the coefficients
\begin{equation}
a_{\mu+1} \ = \ \frac{1}{i2(\mu+1)}
         \left(\mu(\mu+1)-\frac{d-1}{2}\left(\frac{d-1}{2}-1\right)\right)
         a_\mu \ ,
\label{eq1.15}
\end{equation}
with the starting value $a_0=1$. A special situation occurs when the
integer running index $\mu$ reaches the critical value
$\mu_c=(d-3)/2$. If $d$ is odd, $\mu_c$ is integer and the recursion
terminates at $\mu=\mu_c$, i.e.~one has $a_\mu=0, \ \mu>\mu_c$. Thus,
the asymptotic series becomes a {\it finite} polynomial in inverse
powers of $kr$.  However, if $d$ is even, $\mu_c$ is half--odd integer
and the series cannot terminate, it is always {\it infinite}. This
explains the different structure of the Bessel functions in even and
odd dimensional spaces from the viewpoint of the differential
equation.

In App.~\ref{appVBF} we discuss an alternative integral representation
which has an interesting analogue in the matrix space.

\section{Matrix Bessel Functions}
\label{sec2}

We compile the basics features of the matrix spaces we want to work
with in Sec.~\ref{sec2.1}, before we define the matrix Bessel
functions as group integrals in Sec.~\ref{sec2.2}. 

Two general aspects are shifted into the appendix. First, we present
an interesting alternative integral representation in
App.~\ref{appMBF}.  Second, the matrix Bessel functions play a crucial
r\^ole in harmonic analysis or, equivalently, in Fourier--Bessel
analysis in matrix spaces. For the general theory, we refer the reader
to Harish--Chandra's treatise in Ref.~\cite{HC} and to 
Helgason's book~\cite{HEL}. However, to achieve
our goal of being explicit, we collect, for the convenience of the
reader, some results for the Fourier--Bessel analysis of invariant
functions in matrix spaces in App.~\ref{appFBA}.

\subsection{Basics and Notation}
\label{sec2.1}

We introduce $N\times N$ matrices $H$ whose elements $H_{nm}, \ 
n,m=1,\ldots,N$ are real, complex or quaternion variables. In other
words, each element $H_{nm}$ has $\beta$ real components
$H_{nm}^{(\alpha)}, \ \alpha=0,\ldots,(\beta-1)$ with $\beta=1,2,4$,
respectively,
\begin{equation}
H_{nm} \ = \ \sum_{\alpha=0}^{\beta-1} H_{nm}^{(\alpha)} \, 
                                 \tau^{(\alpha)} \ .
\label{eq1.21}
\end{equation}
Here, we use the basis $\tau^{(\alpha)}, \ \alpha=0,\ldots,(\beta-1)$.
We have $\tau^{(0)}=1$ for the real case with $\beta=1$.  For the
complex case with $\beta=2$, we have $\tau^{(0)}=1$ and
$\tau^{(1)}=i$. Finally, we have
\begin{eqnarray}
\tau^{(0)} \ = \ \left[ \begin{array}{cc}
                        1 & 0 \\
                        0 & 1
                        \end{array} \right] \ , \qquad
\tau^{(1)} \ = \ \left[ \begin{array}{cc}
                         0 & +1 \\
                        -1 &  0
                        \end{array} \right] \ , \nonumber\\
\tau^{(2)} \ = \ \left[ \begin{array}{cc}
                         0 & -i \\
                        -i &  0
                        \end{array} \right] \ , \qquad
\tau^{(3)} \ = \ \left[ \begin{array}{cc}
                        +i &  0 \\
                         0 & -i
                        \end{array} \right] \ .
\label{eq1.22}
\end{eqnarray}
in the quaternion case for $\beta=4$ where the $\tau^{(\alpha)}, \ 
\alpha=1,2,3$ are the Pauli matrices. We notice that the total $H$ is
a $2N \times 2N$ matrix for $\beta=4$. However, here and in the
following, the dimensions that we use always refer to the number of
matrix elements such as $H_{nm}$. These are scalar for $\beta=1,2$ and
quaternion for $\beta=4$.  The label $\beta$ is often referred to as
Dyson index.

We assume that the matrix $H$ is real symmetric, Hermitean or
Hermitean self--dual in the three cases $\beta=1,2,4$.  We always
write $H^\dagger=H$ to indicate this symmetry.  There are $N$
independent real variables $H_{nn}=H_{nn}^{(0)}, \ n=1,\ldots,N$ on
the diagonal and $\beta N(N-1)/2$ independent real variables
$H_{nm}^{(\alpha)}, \ \alpha=0,\ldots,(\beta-1), \ 1 \le n < m \le N$
outside the diagonal.  We write the volume element of $H$ in the form
\begin{equation}
d[H] \ = \ \prod_{n=1}^N dH_{nn}^{(0)} 
           \prod_{n<m} \prod_{\alpha=0}^{\beta-1} dH_{nm}^{(\alpha)} 
\label{eq1.22a}
\end{equation}
The matrix $H$ is diagonalized by the matrix $U$, with columns $U_n, \ 
n=1,\ldots,U_N$. Depending on the value of $\beta$, the matrix $U$ is
either orthogonal, unitary or unitary--symplectic. Following Gilmore's
notation~\cite{Gil}, we write $U \in U(N;\beta)$ with $U(N;1)=SO(N)$,
$U(N;2)=U(N)$ and $U(N;4)=USp(2N)$. The volume of these groups is
given by
\begin{equation}
{\rm vol\,} U(N;\beta) \ = \ \prod_{n=1}^N 
                    \frac{2\pi^{\beta n/2}}{\Gamma(\beta n/2)} 
                       \ = \ \frac{2^N\pi^{\beta N(N+1)/4}}
                             {\prod_{n=1}^N \Gamma(\beta n/2)} \ .
\label{eq1.22b}
\end{equation}
We use it to normalize the invariant measure $d\mu(U)$ of 
$U \in U(N;\beta)$ to
unity,
\begin{equation}
\int d\mu(U) \ = \ 1 \ .
\label{eq1.22c}
\end{equation}
The $N$ real eigenvalues $x_n, \ n=1,\dots,N$ of $H$ are ordered in
the diagonal matrix $x$. We have $x=\diag(x_1,\ldots,x_N)$ for
$\beta=1$ and $\beta=2$.  For $\beta=4$, the eigenvalues are doubly
degenerate and we have $x=\diag(x_1,x_1,\ldots,x_N,x_N)$.  Physically,
this doubling of the eigenvalues is due to Kramer's degeneracies.
Thus, the diagonalization reads
\begin{equation}
H \ = \ U^\dagger x U \ , \qquad  {\rm with} \qquad
H_{nm} \ = \ U_n^\dagger x U_m \ .
\label{eq1.23}
\end{equation}
The diagonalizing matrix $U$ has the property $U^{-1}=U^\dagger$.  The
volume element in eigenvalue--angle coordinates is given
by~\cite{Mehta,Hua}
\begin{equation}
d[H] \ = \ C_N^{(\beta)} |\Delta_N(x)|^\beta d[x] d\mu(U)
\label{eq1.23a}
\end{equation}
where $d[x]$ denotes the product of all differentials $dx_n$.  We have
introduced the Vandermonde determinant
\begin{equation}
\Delta_N(x) \ = \ \prod_{n<m} (x_n-x_m) \ .
\label{eq1.23b}
\end{equation}
The normalization constant 
\begin{equation}
C_N^{(\beta)} \ = \ \frac{\pi^{\beta N(N-1)/4}}{N!} 
                    \frac{\Gamma^N(\beta/2)}
                         {\prod_{n=1}^N\Gamma(\beta n/2)}
\label{eq1.23c}
\end{equation}
obtains from the constants given in Mehta's book~\cite{Mehta} and from
Eq.~(\ref{eq1.22b}).

To avoid inconveniences and to ensure a compact notation, we define
the trace $\Tr$ and the determinant $\Det$ with $\Tr=\tr$ and
$\Det=\det$ for $\beta=1,2$ and with
\begin{equation}
\Tr K \ = \ \frac{1}{2} \tr K \qquad {\rm and} \qquad
\Det K \ = \ \sqrt{\det K}
\label{eq1.24}
\end{equation}
in the case $\beta=4$ for a matrix $K$ with quaternion entries.  If
$k$ denotes the diagonal matrix of the eigenvalues of a
real--symmetric, Hermitean or Hermitean self--dual matrix, it is also
useful to define the associate matrix $\hat{k}$. In all three cases
$\beta$, it is the $N \times N$ matrix
$\hat{k}=\diag(k_1,k_2,\ldots,k_N)$, i.e.~we have $\hat{k}=k$ for
$\beta=1,2$ and no degeneracies for $\beta=4$.

\subsection{Integral Definition and Differential Equation}
\label{sec2.2}

As in the case of vector Bessel functions, we start in the matrix case
with the plane wave. For two matrices $H$ and $K$ with the same
symmetries $H^\dagger=H$ and $K^\dagger=K$, we introduce the matrix plane
wave as $\exp\left(i\Tr HK\right)$ where the trace is the proper scalar
product in the matrix space. The matrix plane wave has the property
\begin{equation}
\frac{1}{(2\pi)^N\pi^{\beta N(N-1)/2}} 
 \int d[H] \exp\left(i\Tr HK\right) \ = \ \delta(K) 
\label{eq1.24a}
\end{equation}
where $\delta(K)$ is the product of the $\delta$ distributions of all
independent variables.  We define the matrix gradient
$\partial/\partial H$ and the Laplacean operator
\begin{equation}
\Delta \ = \ \Tr\frac{\partial^2}{\partial H^2} 
       \ = \ \sum_{n=1}^N\frac{\partial^2}{\partial H_{nn}^{(0)2}} 
             \, + \, \frac{1}{2}
             \sum_{n<m} \sum_{\alpha=0}^{\beta-1}
                  \frac{\partial^2}{\partial H_{nm}^{(\alpha)2}} 
\label{eq1.25}
\end{equation}
which acts on the matrix plane wave as
\begin{equation}
\Delta \exp\left(i\Tr HK\right) \ = \ 
         -\Tr K^2 \exp\left(i\Tr HK\right) \ .
\label{eq1.26}
\end{equation}
We notice that, for $\beta=4$, inconvenient factors of two would occur
if we used $\tr$ instead of $\Tr$.

Analogously to vector Bessel functions, we define the matrix Bessel
functions as the angular average
\begin{equation}
\Phi_N^{(\beta)}(x,k) \ = \ \int d\mu(U) \exp(i\Tr HK) \ .
\label{eq1.27}
\end{equation}
The diagonal matrix $k$ contains the eigenvalues of $K$ which is
diagonalized by a matrix $V$ such that $K=V^\dagger kV$. Due to the
invariance of the measure $d\mu(U)$, the matrix $V$ is absorbed and
the functions $\Phi_N^{(\beta)}(x,k)$ depend on the eigenvalues $x$
and $k$ only,
\begin{equation}
\Phi_N^{(\beta)}(x,k) \ = \ \int d\mu(U) \exp(i\Tr U^\dagger xUk) \ .
\label{eq1.27a}
\end{equation}
Thus, in the scalar product $\Tr HK$, solely the
relative angles between $H$ and $K$ matter. The matrix Bessel
functions are symmetric in the arguments,
\begin{equation}
\Phi_N^{(\beta)}(x,k) \ = \ \Phi_N^{(\beta)}(k,x) 
\label{eq1.27b}
\end{equation}
and normalized to unity,
\begin{equation}
\Phi_N^{(\beta)}(x,0) \ = \ 1 \qquad {\rm and} \qquad
\Phi_N^{(\beta)}(0,k) \ = \ 1 \ .
\label{eq1.27c}
\end{equation}
due to Eq.~(\ref{eq1.22c}). These are spherical functions in the 
sense of Ref.~\cite{GEL1}.

As in the vector case, the differential equation is obtained by
averaging Eq.~(\ref{eq1.26}) over the relative angles,
\begin{equation}
\Delta \int d\mu(V) \exp\left(i\Tr HK\right) \ = \
         - \Tr K^2 \, \int d\mu(V)
              \exp\left(i\Tr HK\right) \ .
\label{eq1.28}
\end{equation}
Again, the Laplacean $\Delta$ commutes with the integral,
because the former is in the space of the matrix $H$, the latter
over the diagonalizing matrix $V$ of $K$. The integral also
commutes with $\Tr K^2=\Tr k^2$. Due to the symmetry between $H$
and $K$, the integral is obviously identical to the 
definition~(\ref{eq1.27a}) and we find
\begin{equation}
\Delta_x \Phi_N^{(\beta)}(x,k) \ = \ 
         - \Tr k^2 \, \Phi_N^{(\beta)}(x,k) \ . 
\label{eq1.29}
\end{equation}
Since the matrix Bessel function $\Phi_N^{(\beta)}(x,k)$ depends only
on the radial variables, i.e.~on the eigenvalues, we replaced the full
Laplacean with its radial part $\Delta_x$. Because of the
transformation rule~(\ref{eq1.23a}), it reads
\begin{eqnarray}
\Delta_x &=& \sum_{n=1}^N \frac{1}{|\Delta_N(x)|^\beta}
               \frac{\partial}{\partial x_n} |\Delta_N(x)|^\beta
               \frac{\partial}{\partial x_n} \nonumber\\ 
         &=& \sum_{n=1}^N \frac{\partial^2}{\partial x_n^2}
               \, + \, \sum_{n<m} \frac{\beta}{x_n-x_m} 
               \left(\frac{\partial}{\partial x_n}-
                     \frac{\partial}{\partial x_m}\right) 
\label{eq1.30}
\end{eqnarray}
We notice that these steps are fully parallel to the corresponding
discussion in Sec.~\ref{sec1}. Importantly, due to the
symmetry~(\ref{eq1.27b}), the functions $\Phi_N^{(\beta)}(x,k)$ must
also solve the differential equation in the $k_n, \ n=1,\ldots,N$
which results from Eq.~(\ref{eq1.29}) by exchanging $x$ and $k$.
Obviously, this a very restrictive requirement.

Comparing the radial operator~(\ref{eq1.10}) in the vector case and
the radial operator~(\ref{eq1.30}), we see that it is the $\beta$ that
corresponds to the {\it spatial} dimension $d$ or, more precisely, to
$d-1$.  The r\^ole played by the {\it matrix} dimension $N$ is a
different one. To illustrate this, we study the two simplest cases.
First, we can formally set $N=1$ and find from the
definition~(\ref{eq1.27}) that
$\Phi_1^{(\beta)}(x,k)=\exp\left(ix_1k_1\right)$ where $H_{11}=x_1$
and $K_{11}=k_1$. In this case, the matrix $U$ has dropped out
trivially. This reflects simply that the scalar product
$\vec{k}\cdot\vec{r}$ is linear in the relative solid angle $\Omega$
between the vectors whereas the scalar product $\Tr HK$ is quadratic
in the relative diagonalizing matrix $U$. The corresponding radial
Laplacean $\Delta_x$ for $N=1$ is identical to the Cartesean $\Delta$.
Therefore, the case $N=1$ is too trivial to give any further insight.
Second, we set $N=2$ and find straightforwardly from the differential
equation~(\ref{eq1.29})
\begin{equation}
\Phi_2^{(\beta)}(x,k) \ = \ 
   \exp\left(i\frac{(x_1+x_2)(k_1+k_2)}{2}\right) \,
   \chi^{(\beta+1)}\left(\frac{(x_1-x_2)(k_1-k_2)}{2}\right) 
\label{eq1.31}
\end{equation}
where $\chi^{(d)}$ is the vector Bessel function in $d$ dimensions as
defined in Eq.~(\ref{eq1.3}). This functions appears in the solution,
because the differences $x_1-x_2$ and $k_1-k_2$ directly correspond to
the lengths $|\vec{r}|$ and $|\vec{k}|$. In higher matrix dimensions
$N$, this simple correspondence is lost. However, we will see in great
detail that the features of the functions $\Phi_N^{(\beta)}(x,k)$, in
particular whether or not explicit solutions can be constructed, are
stronger influenced by $\beta$ than by $N$.

Another point in this context deserves to be underlined.  In the
vector case, the differential equation~(\ref{eq1.10}) and the
solution~(\ref{eq1.6}) were constructed for integer dimensions $d$.
However, both equations are also well defined for any real and
positive $d$. Similarly, we observe in the matrix case that the
differential equation~(\ref{eq1.29}) was derived for the cases
$\beta=1,2,4$. However, neither itself nor its solution~(\ref{eq1.31})
for $N=2$ are confined to these cases $\beta=1,2,4$, they are valid
for any real and positive $\beta$. Thus, the cases $\beta=1,2,4$ which
correspond to a matrix model, i.e.~to the defining
integral~(\ref{eq1.27}) of the matrix Bessel functions, are only
special cases of a much more general problem, namely finding the
solutions of the differential equation~(\ref{eq1.29}) for every
integer $N$ and for {\it arbitrary} real values of $\beta$.
We will return to this in Sec.~\ref{sec7}.

\section{Recursion Formula}
\label{sec4}

The matrix Bessel functions show a recursive structure which we
construct by introducing radial Gelfand--Tzetlin coordinates. 
The result is stated in Sec.~\ref{sec4.1} and derived
in Sec.~\ref{sec4.2}. The corresponding invariant measure is 
calculated in Sec.~\ref{sec4.3}.

\subsection{Statement of the Result}
\label{sec4.1}

The matrix Bessel functions, defined in Eq.~(\ref{eq1.27a}),
\begin{equation}
\Phi_N^{(\beta)}(x,k) \ = \ \int d\mu(U) \, \exp(i\Tr U^\dagger xUk)
\label{eq2.1}
\end{equation}
depend on the radial space of the eigenvalues $x$ and $k$. As before,
we write $x=\diag(x_1,\ldots,x_N)$ and $k=\diag(k_1,\ldots,k_N)$ for
$\beta=1,2$ and, for $\beta=4$, we write $x=\diag(x_1,x_1,\ldots,x_N,
x_N)$ and $k=\diag(k_1,k_1,\ldots,k_N,k_N)$. We emphasize that the
radial spaces do not lie in the manifolds covered by the groups
$U(N;\beta)$.  However, we will show that the group
integral~(\ref{eq2.1}) can be exactly mapped onto a recursive
structure which acts exclusively in the radial space. This remarkable
feature is the main result of this section.

Under rather general circumstances, the matrix Bessel functions
$\Phi_N^{(\beta)}(x,k)$ can be calculated iteratively by the explicit
{\it recursion formula}
\begin{equation}
\Phi_N^{(\beta)}(x,k) \ = \ 
         \int d\mu(x^\prime,x) \, 
         \exp\left(i(\Tr x - \Tr x^\prime)k_N \right) \, 
         \Phi_{N-1}^{(\beta)}(x^\prime,\widetilde{k})
\label{eq2.2}
\end{equation}
where $\Phi_{N-1}^{(\beta)}(x^\prime,\widetilde{k})$ is the group
integral (\ref{eq2.1}) over $U(N-1;\beta)$. We have introduced the
diagonal matrix $\widetilde{k}=\diag(k_1,\ldots,k_{N-1})$ for
$\beta=1,2$ and $\widetilde{k}=\diag(k_1,k_1,\ldots,k_{N-1},k_{N-1})$
for $\beta=4$ such that $k=\diag(\widetilde{k},k_N)$ for $\beta=1,2$
and $k=\diag(\widetilde{k},k_N,k_N)$ for $\beta=4$.  Importantly, the
$N-1$ integration variables $x_n^\prime, \ n=1,\ldots,N-1$, ordered in
the diagonal matrix $x^\prime=\diag(x_1^\prime,\ldots,x_{N-1}^\prime)$
for $\beta=1,2$ and $x^\prime=\diag(x_1^\prime,x_1^\prime,\ldots,
x_{N-1}^\prime,x_{N-1}^\prime)$ for $\beta=4$ are arguments of
$\Phi_{N-1}^{(\beta)}(x^\prime,\widetilde{k})$. Moreover, we notice
that their further appearance in the exponential is a simple one due
to the trace.  

The coordinates $x^\prime$ are constructed in the spirit of, but they
are different from, the Gelfand--Tzetlin coordinates of
Refs.~\cite{GT,SLS}.  To clearly distinguish these two sets of
coordinates from each other, we refer to the latter as {\it angular}
Gelfand--Tzetlin coordinates and to the variables $x^\prime$ as {\it
  radial} Gelfand--Tzetlin coordinates.  The difference is at first
sight minor, but of crucial importance.  In the angular case, $x$ is
in the Cartan subalgebra belonging to $U(N;\beta)$.  In the radial
case, however, $x$ is in the {\it radial} space of the eigenvalues of
the real--symmetric, Hermitean or Hermitean self--dual matrix $H$,
which are the arguments of the functions (\ref{eq2.1}).
While the angular Gelfand--Tzetlin coordinates never leave the group
space, the radial ones establish an exact and unique relation between
the group and the radial space.  The radial Gelfand--Tzetlin
coordinates re--parametrize the sphere that is described by the
$N^{\rm th}$ column $U_N$ of the matrix $U \in U(N;\beta)$.  The
recursion formula~(\ref{eq2.2}) can only be constructed in the radial
coordinates $x^\prime$, but not in the angular ones.  The radial and
the angular Gelfand--Tzetlin coordinates are, in general, different.
They happen to coincide for $\beta=2$, i.e.~for the unitary group
$U(N)$. This illustrates, in the framework of our recursion formula,
the special r\^ole played by the unitary group.

The invariant measure $d\mu(x^\prime,x)$ is, apart from phase angles,
the invariant measure $d\mu(U_N)$ on the sphere in question, expressed
in the radial coordinates $x^\prime$. It only contains algebraic
functions and reads explicitly
\begin{equation}
d\mu(x^\prime,x) \ = \ \frac{2^{N-1}\Gamma(N\beta/2)}
                            {\pi^{N(\beta-2)(\beta-4)/6}} \,
                       \frac{\Delta_{N-1}(x^\prime)}
                            {\Delta_N^{\beta-1}(x)} \,
                        \left(-\prod_{n,m}(x_n-x_m^\prime)\right)
                                                      ^{(\beta-2)/2} \,
                        d[x^\prime] \ .
\label{eq2.2a}
\end{equation}
The normalization constant obtains from results in Gilmore's
book~\cite{Gil}. It ensures normalization to unity according to
Eq.~(\ref{eq1.22c}). The domain of integration is compact and given by
\begin{equation}
x_n \ \le \ x_n^\prime \ \le \ x_{n+1} \ , \qquad n=1,\ldots,(N-1) \ ,
\label{eq2.2b}
\end{equation}
reflecting a ``betweenness condition'' for the radial Gelfand--Tzetlin
coordinates. This is why no absolute value signs appear in the
measure~(\ref{eq2.2a}).

The general recursion formula~(\ref{eq2.2}) states an iterative way
for constructing the matrix Bessel function $\Phi_N^{(\beta)}(x,k)$
for arbitrary $N$ from the matrix Bessel function
$\Phi_2^{(\beta)}(x,k)$ for $N=2$ which can usually be obtained
trivially.  We remark that the recursion formula allows one to express
the matrix Bessel functions in the form
\begin{eqnarray}
\Phi_N^{(\beta)}(x,k) &=&
       \int \prod_{n=1}^{N-1} d\mu(x^{(n)},x^{(n-1)}) 
                             \nonumber\\
    & & \qquad \exp\left(i(\Tr x^{(n-1)}-\Tr x^{(n)})k_{N-n+1}\right)
        \, \exp\left(ix_1^{(N-1)}k_1\right)
\label{eq2.3a}
\end{eqnarray}
where we have introduced the radial Gelfand--Tzetlin coordinates
$x_m^{(n)}, \ m=1,\ldots,N-n$ on $N-1$ levels $n=1,\ldots,(N-1)$.  We
define $x^{(0)}=x$ and $x^{(1)}=x^\prime$.

\subsection{Derivation}
\label{sec4.2}

We introduce a matrix $V=\diag(\widetilde{V},V_0)$ with
$\widetilde{V}\in U(N-1;\beta)$ and $V_0\in U(1;\beta)$ such that
$V\in U(N-1;\beta)\otimes U(1;\beta) \subset U(N;\beta)$ and multiply the
right hand side of the definition~(\ref{eq2.1}) with
\begin{equation}
1 \ = \ \int d\mu(V) 
  \ = \ \int d\mu(V_0) \, \int d\mu(\widetilde{V}) \ .
\label{eq2.5}
\end{equation}
The invariance of the Haar measure $d\mu(U)$ allows us to replace
$U$ with $UV^\dagger$ and to write
\begin{equation}
\Phi_N^{(\beta)}(x,k) \ = \ \int d\mu(V_0) \int d\mu(\widetilde{V}) 
                  \int d\mu(U) \, \exp(i\Tr U^\dagger xUV^\dagger k V)
                  \ .
\label{eq2.6}
\end{equation}
We collect the first $N-1$ columns $U_n$ of $U$ in the $N\times(N-1)$
rectangular matrix $B$ such that $B=[U_1\, U_2\, \cdots U_{N-1}]$ and
$U=[B\, U_N]$. We notice that
\begin{eqnarray}
B^\dagger B &=& 1_{N-1} \nonumber\\
B B^\dagger &=& \sum_{n=1}^{N-1} U_nU_n^\dagger 
           \ = \ 1_N-U_NU_N^\dagger \ . 
\label{eq2.6a}
\end{eqnarray}
As already stated in Sec.~\ref{sec2.1}, the elements of a vector or a
matrix are scalar for $\beta=1,2$ and quaternion for $\beta=4$.  In
this sense, we also write $1_N$ as the unit matrix for $\beta=4$
because its elements are $\tau^{(0)}$.  By defining the
$(N-1)\times(N-1)$ square matrices $\widetilde{H}=B^\dagger x B$ and
$\widetilde{K}=\widetilde{V}^\dagger \widetilde{k} \widetilde{V}$ we
may rewrite the trace in Eq.~(\ref{eq2.6}) as
\begin{equation}
\Tr U^\dagger xUV^\dagger k V \ = \ \Tr\widetilde{H}\widetilde{K}
                                   \, + \, H_{NN} k_N
\label{eq2.7}
\end{equation}
with $H_{NN}=U_N^\dagger xU_N$ according to Eq.~(\ref{eq1.23}).  We
notice that $V_0$ has dropped out.  Since the first term of the right
hand side of Eq.~(\ref{eq2.7}) depends only on the first $N-1$ columns
$U_n$ collected in $B$ and the second term depends only on $U_N$, we
use the decomposition
\begin{equation}
d\mu(U) \ = \ d\mu(B) \, d\mu(U_N)
\label{eq2.8}
\end{equation}
of the measure to cast Eq.~(\ref{eq2.6}) into the form
\begin{equation}
\Phi_N^{(\beta)}(x,k) \ = \ \int d\mu(U_N) \, \exp(iH_{NN} k_N) \,
                  \int d\mu(\widetilde{V}) \int d\mu(B)
                  \, \exp(i\Tr\widetilde{H}\widetilde{K})
\label{eq2.9}
\end{equation}
where we have already done the trivial integration over $V_0$.  

The difficulty to overcome lies in the decomposition~(\ref{eq2.8}).
While $d\mu(U_N)$ is simply the invariant measure on the sphere
described by $U_N$, the measure $d\mu(B)$ is rather complicated.
Pictorially speaking, the degrees of freedom in $d\mu(B)$ have always
to know that they are locally orthogonal to $U_N$. Thus, $d\mu(B)$
depends on $U_N$.  Luckily, there is one distinct set of coordinates
that is perfectly suited to this situation.  It is the system of the
radial Gelfand--Tzetlin coordinates. We construct it by transferring
the methods of Ref.~\cite{GGT} for the angular case to the radial
case.

The $N\times N$ matrix $(1_N-U_NU_N^\dagger)$ is a projector onto the
$(N-1)\times (N-1)$ space obtained from the original $N\times N$ space
by slicing off the vector $U_N$. We project the radial coordinates
$x$ onto this space and study its spectrum. The defining equation reads
\begin{equation}
(1_N-U_NU_N^\dagger) \, x \, (1_N-U_NU_N^\dagger) \, E_n^\prime \ = \
                            x_n^\prime \, E_n^\prime 
\qquad n=1,\ldots,N-1 \ .
\label{eq2.10}
\end{equation}
Equation~(\ref{eq2.10}) determines the $N-1$ radial Gelfand--Tzetlin
coordinates $x_n^\prime$ and the corresponding vectors $E_n^\prime$ as
eigenvalues and eigenvectors of the matrix $(1_N-U_NU_N^\dagger)\, x
\, (1_N-U_NU_N^\dagger)$ which has the rank $N-1$.  Since we have by
construction $U_N^\dagger E_n^\prime = 0$, we may as well write
\begin{equation}
(1_N-U_NU_N^\dagger) \, x \, E_n^\prime \ = \
                            x_n^\prime \, E_n^\prime \ ,
\qquad n=1,\ldots,N-1 \ .
\label{eq2.11}
\end{equation}
The eigenvalues $x_n^\prime, \ n=1,\ldots,N$ are obtained from the
characteristic equation
\begin{eqnarray}
0 &=& \Det\left((1_N-U_NU_N^\dagger)x-x_n^\prime\right)
                               \nonumber\\
  &=& \Det\left(x-x_n^\prime\right) \, 
      \det\left(1_N-(x-x_n^\prime)^{-1}U_NU_N^\dagger x\right)
                               \nonumber\\
  &=& \Det\left(x-x_n^\prime\right) \, 
      \left(1-U_N^\dagger\frac{x}{x-x_n^\prime}U_N\right)
                               \nonumber\\
  &=& -x_n^\prime \, \Det\left(x-x_n^\prime\right) \,
        \Tr U_N^\dagger\frac{1_N}{x-x_n^\prime}U_N \ .
\label{eq2.12}
\end{eqnarray}
Together with the normalization $\Tr U_N^\dagger U_N=1$,
this yields the $N$ equations
\begin{eqnarray}
1 &=& \Tr U_N^\dagger U_N 
      \ = \ \sum_{n=1}^N \sum_{\alpha=0}^{\beta-1} U_{nN}^{(\alpha)2}
                                      \nonumber\\ 
0 &=& \Tr U_N^\dagger \frac{1_N}{x-x_n^\prime} U_N 
      \ = \ \sum_{m=1}^N \sum_{\alpha=0}^{\beta-1}
               \frac{U_{mN}^{(\alpha)2}}{x_m-x_n^\prime} \ , 
                               \qquad n=1,\ldots,N-1 \ .
\label{eq2.3}
\end{eqnarray}
In these formulae, the trace $\Tr$ is only needed in the symplectic
case. We notice that the equations for the variables $x^\prime$ depend
on the variables $x$ as parameters.  We emphasize once more that $x$
in these equations is in the radial space and, in general, not in the
Cartan subalgebra of $U(N;\beta)$.

At this point, it is not clear yet why the introduction of the radial
Gelfand--Tzetlin coordinates is at all helpful. The great advantage
will reveal itself when we express the matrix $\widetilde{H}$ and the
matrix element $H_{NN}$ in the trace~(\ref{eq2.9}) in these coordinates.
To this end, we first multiply Eq.~(\ref{eq2.10}) from the right
with $E_n^{\prime\dagger}$ and sum over $n$,
\begin{equation}
(1_N-U_NU_N^\dagger) \, x \, (1_N-U_NU_N^\dagger) \ = \
            \sum_{n=1}^{N-1} x_n^\prime \,
            E_n^\prime E_n^{\prime\dagger} 
\label{eq2.13}
\end{equation}
where we used the completeness relation
\begin{equation}
\sum_{n=1}^{N-1} E_n^\prime E_n^{\prime\dagger} 
       \, + \, U_NU_N^\dagger \ = \ 1_N \ .
\label{eq2.13a}
\end{equation}
Taking the trace of the spectral expansion~(\ref{eq2.13}) we find
immediately
\begin{equation}
\Tr x \, - \, \Tr x^\prime \ = \ \Tr U_N^\dagger x U_N \ = \ H_{NN} \ 
\label{eq2.14}
\end{equation}
This is a remarkably simple result.  An analogous expression exists
for the $NN$ matrix element of the unitary group in the theory of
angular Gelfand--Tzetlin coordinates for the unitary
group~\cite{GT,SLS}.  Here we have shown that Eq.~(\ref{eq2.14}) is a
general feature in every radial space.

We now turn to the $(N-1)\times(N-1)$ matrix $\widetilde{H}$. Its
$N-1$ eigenvalues $y_n, \ n=1,\ldots,N-1$ are determined by the
characteristic equation
\begin{eqnarray}
0 \ = \ \Det\left(\widetilde{H}-y_n\right)
  &=& \Det\left(B^\dagger xB-y_n\right)  
                               \nonumber\\
  &=& - \frac{1}{y_n} \, \Det\left(BB^\dagger x-y_n\right)  
                               \nonumber\\
  &=& - \frac{1}{y_n} \, \Det\left((1_N-U_NU_N^\dagger)x-y_n\right) 
\label{eq2.15}
\end{eqnarray}
where we used Eq.~(\ref{eq2.6a}) and re--expressed a $(N-1)\times
(N-1)$ determinant as a $N\times N$ determinant. The comparison
of Eq.~(\ref{eq2.15}) with Eq.~(\ref{eq2.12}) shows that, most
advantageously, we have $y_n\equiv x_n^\prime, \ n=1,\ldots,N-1$.
Thus we may write
\begin{equation}
\widetilde{H} \ = \ B^\dagger x B 
              \ = \ \widetilde{U}^\dagger x^\prime \widetilde{U} 
\label{eq2.16}
\end{equation}
by introducing the $(N-1) \times (N-1)$ {\it square} matrix
$\widetilde{U}$ which diagonalizes $\widetilde{H}$. Obviously,
$\widetilde{U}$ must be a complicated function of the $N \times (N-1)$
{\it rectangular} matrix $B$, i.e.~of the columns $U_n, \ 
n=1,\ldots,N-1$. However, all we need to know is that $\widetilde{U}$
must be in the group $U(N-1;\beta)$ because, by construction,
$\widetilde{H}$ has the symmetry
$\widetilde{H}^\dagger=\widetilde{H}$.

Collecting everything, we cast Eq.~(\ref{eq2.9}) into the form
\begin{eqnarray}
\Phi_N^{(\beta)}(x,k) &=& \int d\mu(x^\prime,x) \, 
                  \exp\left(i(\Tr x - \Tr x^\prime) k_N\right) 
                      \nonumber\\
            & & \qquad \int d\mu(\widetilde{V}) \int d\mu(B)
                  \, \exp(i\Tr\widetilde{U}^\dagger x^\prime \widetilde{U}
            \widetilde{V}^\dagger \widetilde{k} \widetilde{V}) \ .
\label{eq2.17}
\end{eqnarray}
We may now use the invariance of the Haar measure
$d\mu(\widetilde{V})$ to absorb $\widetilde{U}$ such that
\begin{eqnarray}
\Phi_N^{(\beta)}(x,k) &=& \int d\mu(x^\prime,x) \, 
                  \exp\left(i(\Tr x - \Tr x^\prime) k_N\right) 
                      \nonumber\\
            & & \qquad \int d\mu(\widetilde{V}) 
    \, \exp(i\Tr x^\prime \widetilde{V}^\dagger\widetilde{k}\widetilde{V}) 
                       \int d\mu(B) \ .
\label{eq2.18}
\end{eqnarray}
Thus, the integration over $B$ is trivial and yields unity due to our
normalization. The remaining integration over $\widetilde{V}$ gives
precisely the matrix Bessel function
$\Phi_{N-1}^{(\beta)}(x^\prime,\widetilde{k})$.  This completes the
derivation of the recursion formula in Sec.~\ref{sec4.1}.  The reader
experienced with group integration has realized that the introduction
of the matrix $V=\diag(\widetilde{V},V_0)$ was not strictly necessary.
Alternatively, one could have shown that the measure $d\mu(B)$ can be
identified with $d\mu(\widetilde{U})$ and have done the corresponding
integral. However, we believe that the introduction of $V$ makes this
part of the derivation more transparent.

\subsection{Invariant Measure}
\label{sec4.3}

The invariant measure $d\mu(U_N)$ has to be expressed in terms of the
radial coordinates $x^\prime$.  To this end, we first have to solve
Eq.~(\ref{eq2.3}) for the moduli squared of the vector $U_N$ as a
function of the new coordinates $x^\prime$.  Since Eq.~(\ref{eq2.3})
for the total moduli square for all $\beta$ coincides with the
equation for the {\it angular} Gelfand--Tzetlin coordinates of the
unitary group, we can use the results as derived in~\cite{GT,SLS}.  We
have in the three cases
\begin{equation}
|U_{nN}|^2 \ = \
\sum_{\alpha=0}^{\beta-1} (U_{nN}^{(\alpha)})^2 \ = \
     \frac{\prod_{m=1}^{N-1}(x_n-x_m^\prime)}
     {\prod_{m\neq n}(x_n-x_m)} \ .
\label{inv1}
\end{equation}
The betweenness condition~(\ref{eq2.2b}) follows from the positive
definiteness of this expression. We parametrize the remaining degrees
of freedom of $U_{nN}$ in the cases $\beta=2,4$. We set $U_{nN}^{(0)}
= \cos \gamma_n$ and $U_{nN}^{(1)} = \sin \gamma_n$ in the case
$\beta=2$ and
\begin{equation}
U_{nN}=\left[
       \matrix{\cos \psi_n\exp(i \gamma_n^{(1)})& 
               \sin \psi_n\exp(i \gamma_n^{(2)})\cr
               -\sin \psi_n\exp(-i \gamma_n^{(2)})&
               \cos \psi_n\exp(-i \gamma_n^{(1)})}\right]
\label{inv2}
\end{equation}
for $\beta=4$ in the basis (\ref{eq1.22}).  The invariant length
element reads
\begin{eqnarray}
\Tr dU_N^\dagger dU_N &=& \sum_{n=1}^N \sum_{\alpha=0}^{\beta-1} 
                          \left(dU_{nN}^{(\alpha)}\right)^2 
                                    \nonumber\\
                      &=& \sum_{n=1}^N 
                          \biggl(\frac{1}{4|U_{nN}|^2}(d|U_{nN}|^2)^2
                          +\sum_{i=1}^{\beta/2}|U_{nN}|^2(d\gamma_n^{(i)})^2
                                                 \biggr.
                                    \nonumber\\
                      & & \qquad\qquad\qquad\qquad\qquad
                          +\biggl.\delta_{\beta 4}|U_{nN}|^2(d\cos\phi_n)^2
                          \biggr) \ .
\label{inv3}
\end{eqnarray}
To express the differential $d|U_{nN}|^2$ in terms of the
$dx_n^\prime$, we again take advantage of the results in
Refs.~\cite{GT,SLS}
\begin{equation}
\sum_{n=1}^N\frac{1}{4|U_{nN}|^2}(d|U_{nN}|^2)^2 \ = \
   \sum_{n=1}^{N-1}\frac{\prod_{m=1}^{N-1}(x_m^\prime-x_n^\prime)}
   {4\prod_{m=1}^N(x_m-x_n^\prime)}(dx_n^\prime)^2\quad .
\label{inv4}
\end{equation}
{}From these equations, we can read off the metric $g$ in the basis of
the coordinates $x_n^\prime,\gamma_n^{(i)}$ and $\psi_n$. Conveniently,
it is diagonal. The determinant of $g$ is given by 
\begin{equation}
\det g \ = \ \frac{\Delta_{N-1}^2(x^\prime)}
                            {\Delta_N^{2\beta-2}(x)} \,
                           \prod_{n,m}(x_n-x_m^\prime)^{(\beta-2)} \ ,
\label{inv5}
\end{equation}
which yields the invariant measure $d\mu(U_N)$ in terms of the
$x_n^\prime$ and of the additional coordinates $\gamma_n^{(j)}$ and
$\psi_n$. These angles can be integrated out trivially. This yields
Eq.~(\ref{eq2.2a}).

\section{Radial Functions for Arbitrary $\beta$}
\label{sec7}

Remarkably, the recursion introduced in the previous section, is the
iterative solution of the radial equation for arbitrary values of
$\beta$. Thus, the matrix Bessel functions are special cases of more
general functions which we want to refer to as radial functions. We
give the precise formulation of the problem in Sec.~\ref{sec7.1} and
show in Sec.~\ref{sec7.2} that the recursion is the general iterative
solution. In Sec.~\ref{sec7.3}, we discuss a Hankel ansatz for the
radial functions.

\subsection{Definition by the Differential Equation}
\label{sec7.1}

In Sec.~\ref{sec2.2}, we defined the matrix Bessel function through
the group integral~(\ref{eq1.27}) or, equivalently, the group
integral~(\ref{eq1.27a}). This definition confines the dimension
$\beta$ to the values $\beta=1,2,4$, corresponding to the groups
$U(N;\beta)$. However, discussing the simplest case $N=2$, we already
saw in Sec.~\ref{sec2.2} that $\Phi_2^{(\beta)}$ is well defined for
{\it arbitrary} values of $\beta$. This was a simple consequence of
the explicit form~(\ref{eq1.31}) which expresses $\Phi_2^{(\beta)}$ in
terms of the Bessel function $\chi^{(\beta+1)}$. The latter is known
to be well defined for arbitrary $\beta$. Hence, we conclude that
the cases $\beta=1,2,4$ which relate to matrices and groups are 
embedded into a space of far more general functions.

It seems natural that this phenomenon also extends to $N>2$.
The problem has to be posed as follows: We seek the solutions
$\Phi_N^{(\beta)}$ of the differential equation
\begin{equation}
\Delta_x \Phi_N^{(\beta)}(x,k) \ = \ 
         - \sum_{n=1}^N k_n^2 \, \Phi_N^{(\beta)}(x,k)
\label{eq7.1}
\end{equation}
where the operator is given by
\begin{equation}
\Delta_x \ = \ \sum_{n=1}^N \frac{\partial^2}{\partial x_n^2}
               \, + \, \sum_{n<m} \frac{\beta}{x_n-x_m} 
               \left(\frac{\partial}{\partial x_n}-
                     \frac{\partial}{\partial x_m}\right) \ .
\label{eq7.2}
\end{equation}
Here, $\beta$ is arbitrary. For technical reasons, however, we
restrict ourselves for the time being to real and positive values of
$\beta$.  We make no reference whatsoever to matrices, eigenvalues and
groups.  To emphasize this, we view $x$ and $k$ as sets of $N$
variables $x_n, \ n=1,\ldots,N$ and $k_n, \ n=1,\ldots,N$ for every
positive $\beta$.  We do not use traces.

We require that the solutions are symmetric in the argument
\begin{equation}
\Phi_N^{(\beta)}(x,k) \ = \ \Phi_N^{(\beta)}(k,x) 
\label{eq7.3}
\end{equation}
and normalized 
\begin{equation}
\Phi_N^{(\beta)}(0,k) \ = \ 1 \qquad {\rm and} \qquad 
\Phi_N^{(\beta)}(x,0) \ = \ 1
\label{eq7.4}
\end{equation}
at the origin $x=0$ and $k=0$.

In the sequel, we want to refer to the functions
$\Phi_N^{(\beta)}(x,k)$ for arbitrary $\beta$ as {\it radial
  functions} while we reserve the term {\it matrix Bessel functions}
to the cases $\beta=1,2,4$ where the direct connection to matrices and
Lie groups exists.

\subsection{Recursive Solution}
\label{sec7.2}

We claim that the solutions are, for arbitrary $\beta$, given as an
iteration in $N$ by the recursion formula
\begin{equation}
\Phi_N^{(\beta)}(x,k) \ = \ 
         \int d\mu(x^\prime,x) \, 
  \exp\left(i\left(\sum_{n=1}^Nx - 
                  \sum_{n=1}^{N-1}x^\prime\right)k_N\right) \, 
         \Phi_{N-1}^{(\beta)}(x^\prime,\widetilde{k})
\label{eq7.5}
\end{equation}
where $\Phi_{N-1}^{(\beta)}(x^\prime,\widetilde{k})$ is the solution
of the differential equation~(\ref{eq7.2}) for $N-1$.  Here,
$\widetilde{k}$ denotes the set of variables $k_n, \
n=1,\ldots,(N-1)$ and $x^\prime$ the set of integration variables
$x_n^\prime, \ n=1,\ldots,(N-1)$. The integration measure
\begin{equation}
d\mu(x^\prime,x) \ = \ G_N^{(\beta)} \, \frac{\Delta_{N-1}(x^\prime)}
                            {\Delta_N^{\beta-1}(x)} \,
                        \left(-\prod_{n,m}(x_n-x_m^\prime)\right)
                               ^{(\beta-2)/2} \,
                        d[x^\prime] \ .
\label{eq7.6}
\end{equation}
is the continuation of Eq.~(\ref{eq2.2a}) to arbitrary positive
$\beta$.  The normalization constant
\begin{equation}
G_N^{(\beta)} \ = \ 2^{N-1} \, \frac{\Gamma(N\beta/2)}
                                      {\Gamma^N(\beta/2)}
\label{eq7.6a}
\end{equation}
is also the continuation of the constant in Eq.~(\ref{eq2.2a}).  We
calculate it in App.~\ref{appNKB}.  As in the cases $\beta=1,2,4$, the
inequalities
\begin{equation}
x_n \ \le \ x_n^\prime \ \le \ x_{n+1} \ , \qquad n=1,\ldots,(N-1) 
\label{eq7.7}
\end{equation}
define the domain of integration.

We stress that we derived the recursion formula~(\ref{eq7.5}) in
Sec.~\ref{sec4} for the cases $\beta=1,2,4$. To prove that it is the
iterative solution for arbitrary positive $\beta$, we show that it
solves the differential equation~(\ref{eq7.1}). The keystone for the
proof is the identity
\begin{eqnarray}
\Delta_x \Phi_N^{(\beta)}(x,k) 
  &=& -k_N^2 \Phi_N^{(\beta)}(x,k)  
                                          \nonumber\\
  & & \qquad + \, \int d\mu(x^\prime,x) \, 
 \exp\left(i\left(\sum_{n=1}^Nx_n - 
                  \sum_{n=1}^{N-1}x_n^\prime\right)k_N\right) \, 
                                          \nonumber\\
  & & \qquad\qquad\qquad\qquad\qquad
 \Delta_{x^\prime}\Phi_{N-1}^{(\beta)}(x^\prime,\widetilde{k})
\label{eq7.8}
\end{eqnarray}
which is derived in App.~\ref{appLIR}. Equation~(\ref{eq7.8})
establishes a not immediately obvious, but nevertheless natural
connection between, on the one hand, the action of the Laplacean
$\Delta_x$ in the $N$ variables $x_n$ on the radial function in $N$
dimensions, i.e.~on the recursion integral~(\ref{eq7.5}), and, on the
other hand, the recursion integral over the Laplacean
$\Delta_{x^\prime}$ in the $N-1$ variables $x_n^\prime$ acting on the
radial function in $N-1$ dimensions.  There is a compensation term
which is just $-k_N^2 \Phi_N^{(\beta)}(x,k)$. Thus, we can prove the
eigenvalue equation~(\ref{eq7.1}) by induction: assuming that it is
correct for $N-1$, identity~(\ref{eq7.8}) implies Eq.~(\ref{eq7.1})
for $N$. The induction starts with $N=2$ where the eigenvalue
equation~(\ref{eq7.1}) is clearly valid for arbitrary $\beta$ as shown
in Sec.~\ref{sec2.2} by deriving the explicit solution~(\ref{eq1.31}).

The symmetry relation~(\ref{eq7.3}) is non--trivial. In the matrix
cases $\beta=1,2,4$, it is obvious from the integral
definitions~(\ref{eq1.27}) and~(\ref{eq1.27a}). For arbitrary $\beta$,
we cannot use this argument, we only have the recursion~(\ref{eq7.5}).
In App.~\ref{appSYR}, we prove the symmetry relation~(\ref{eq7.3})
by an explicit change of variables.

The normalization $\Phi_N^{(\beta})(x,0)=1$ in Eq.~(\ref{eq7.4})
follows directly from the normalization of the measure~(\ref{eq7.6}).
The symmetry relation~(\ref{eq7.3}) then also yields
$\Phi_N^{(\beta)}(0,k)=1$.

Regarding the domain of $\beta$, a comment is in order. We have
seen in Sec.~\ref{sec2.2} that for $N=2$ the matrix Bessel function
is well defined for arbitrary complex $\beta$. This should also be
true for our recursion formula~(\ref{eq7.5}). However, for $\beta\leq
0$ non--integrable singularities arise at the boundaries in the
integral in Eq.~(\ref{eq7.5}). At the same time the normalization
constant becomes zero for $\beta = 0,-2,-4,\ldots$ compensating the
singularities of the integral. This makes the recursion formula for
$\beta\leq 0$ not ill--defined but it gets more difficult to treat.
Therefore, we have restricted ourselves to positive values of $\beta$.

In the work of Okounkov and Olshanski~\cite{OO} an expansion of the
radial functions for arbitrary $\beta$ in Jack polynomials is derived.
The series run over sets of partitions $\{\mu\}$.  These authors also
derive a recursion formula for the Jack polynomials depending on one
set of continous variables $x$, say, and belonging to such partitions
$\{\mu\}$. It is related to, but different from ours which involves
two sets of continous varables $x$ and $k$. The crucial difference
rests in the exponential function which is present in our
formula~(\ref{eq7.5}), but not in the formula of Ref.~\cite{OO}.
Importantly, it is this exponential term which makes sure that the
symmetry condition~(\ref{eq7.3}) is fulfilled on all levels of the
recursion.  Since the Jack polynomials themselves do not obey such a
symmetry condition, there is no exponential term in the recursion
formula of Ref.~\cite{OO}. However, it must be possible to derive the
recursion formula for the radial functions from the one for the Jack
polynomials. An interesting, although probably not very elegant
approach would be the following: If one inserted the recursion formula
for the Jack polynomials into the expansion~\cite{OO} of the radial
functions in terms of these Jack polynomials, one ought to see that
the series over the partitions can, at least partly, be {\it resummed}
to yield the exponential function present in the recursion
formula~(\ref{eq7.5}). This is remarkable and could be very helpful
for the application of Jack polynomials, because, in general,
resummations over partitions are known to be difficult and involved.
For the connection to Calogero--Sutherland models, we refer the reader
to Sec.~\ref{sec9}.

\subsection{Hankel Ansatz}
\label{sec7.3}

In the spirit of Eq.~(\ref{eq1.12}) for the vector case, we make a
Hankel ansatz for our radial functions for arbitrary positive $\beta$.
We also do this in view of the applications in Sec.~\ref{sec8}.
Since the sum over the $k_n^2$ on the right hand side of the
eigenvalue equation~(\ref{eq7.1}) is invariant under all permutations
of the $k_n$ or, equivalently, their indices $n$, we can label a set
of solutions $\Phi_{N,\omega}^{(\beta)}(x,k)$ by an element $\omega$
of the permutation group $S_N$ of $N$ objects. For these solutions, we
make the ansatz
\begin{equation}
\Phi_{N,\omega}^{(\beta)}(x,k) \ = \ 
       \frac{\exp\left(i\sum_{n=1}^Nx_nk_{\omega(n)}\right)}
                 {|\Delta_N(x)\Delta_N(k)|^{\beta/2}} \,
                 W_{N,\omega}^{(\beta)}(x,k) \ .
\label{eq1.32}
\end{equation}
where $\omega(k)$ is the diagonal matrix constructed from $k$ by
permuting the $k_n$, or the indices $n$. The full solution
$\Phi_N^{(\beta)}(x,k)$, satisfying the constraints~(\ref{eq7.3})
and~(\ref{eq7.4}), is then, apart from possible normalization
constants, given as the linear combination
\begin{equation}
\Phi_{N}^{(\beta)}(x,k) \ = \ \frac{1}{N!}
            \sum_{\omega \in S_N} (-1)^{\pi(\omega)}
                \Phi_{N,\omega}^{(\beta)}(x,k)
\label{eq1.32a}
\end{equation}
of the functions~(\ref{eq1.32}).  Here, $\pi(\omega)$ is the parity of
the permutation.

We find for the function $W_{N,\omega}^{(\beta)}(x,k)$ the
differential equation
\begin{equation}
L_{x,\omega(k)} \, W_{N,\omega}^{(\beta)}(x,k) \ = \ 0
\label{eq1.33}
\end{equation}
where the operator is given by
\begin{equation}
L_{x,\omega(k)} \ = \ 
      \sum_{n=1}^N \frac{\partial^2}{\partial x_n^2} \, + \, 
      i2\sum_{n=1}^N k_{\omega(n)}\frac{\partial}{\partial x_n}
      \, - \, \beta\left(\frac{\beta}{2}-1\right)
              \sum_{n<m} \frac{1}{(x_n-x_m)^2} \ .
\label{eq1.34}
\end{equation}
This differential equation generalizes Eq.~(\ref{eq1.13}) to the
matrix case for $\beta=1,2,4$ and, furthermore, the latter to general
radial function for arbitrary $\beta$.

Again, due to the symmetry~(\ref{eq7.3}), the differential
equation~(\ref{eq1.33}) must also hold if $x$ and $\omega(k)$ are
interchanged. It is the last term of the operator $L_{x,\omega(k)}$
that makes the differential equation~(\ref{eq1.33}) so difficult.
This shows that the case $\beta=2$ corresponding to unitary matrices
$U\in U(N)$ is special: the last term vanishes and we simply have
$W_{N,\omega}^{(2)}(x,k)=1$. This is the Itzykson--Zuber
case~\cite{IZ}. For arbitrary $\beta$, it is obvious from the
differential operator that $W_{N,\omega}^{(\beta)}(x,k) \to 1$ if
$|x_n-x_m|\to\infty$ for all pairs $n<m$. Once more, this must also be
true if $|k_n-k_m|\to\infty$. Thus, we expect that
$W_{N,\omega}^{(\beta)}(x,k)$ is some kind of asymptotic series,
generalizing Eq.~(\ref{eq1.14}) in the vector case.

Hence, we re--derive a known result by concluding that the leading
contribution in an asymptotic expansion of the
functions~(\ref{eq1.32}) is given by
\begin{equation}
\Phi_{N,\omega}^{(\beta)}(x,k) \ \sim \ 
 \frac{\exp\left(i\sum_{n=1}^N x_nk_{\omega(n)}\right)}
      {|\Delta_N(x)\Delta_N(k)|^{\beta/2}} \ .
\label{eq1.34aa}
\end{equation}
According to
Eq.~(\ref{eq1.32a}), this means that
\begin{equation}
\Phi_N^{(\beta)}(x,k) \ \sim \ 
         \frac{\det[\exp(ix_nk_m)]_{n,m=1,\ldots,N}}
              {|\Delta_N(x)\Delta_N(k)|^{\beta/2}} 
\label{eq1.34a}
\end{equation}
is the asymptotic behavior of the radial functions
$\Phi_{N}^{(\beta)}(x,k)$ if the differences $|x_n-x_m|$ and
$|k_n-k_m|$ are large for all pairs $n<m$.

The functions $W_{N,\omega}^{(\beta)}(x,k)$ are translation
invariant, i.e.~they depend only on the differences $(x_n-x_m)$. We
show this in App.~\ref{appWPD}.  Due to the symmetry, this argument
carries also over to $k$ and $W_{N,\omega}^{(\beta)}(x,k)$ depends
only on the differences $(k_n-k_m)$ as well. Moreover, the symmetry
implies that it depends only on the products
$(k_{\omega(n)}-k_{\omega(m)})(x_n-x_m)$.

Collecting all these pieces of information, we make the ansatz
\begin{equation}
W_{N,\omega}^{(\beta)}(x,k) \ = \ \sum_{\{\mu\}}
        \frac{a_{\mu_{12}\mu_{13}\cdots\mu_{(N-1)N}}}
{\prod_{n<m}\left((k_{\omega(n)}-k_{\omega(m)})(x_n-x_m)\right)^{\mu_{nm}}}
\label{eq1.37}
\end{equation}
with coefficients $a_{\mu_{12}\mu_{13}\cdots\mu_{(N-1)N}}$ that depend
on $N(N-1)/2$ integer indices $\mu_{nm}$, as many as there are
differences.  The summation is over the set of these indices. The
presence of the $k_n$ makes it very difficult to solve
Eq.~(\ref{eq1.33}) with the ansatz~(\ref{eq1.37}). In the vector case,
one easily sees that the differential equation~(\ref{eq1.13}) in $r$
can be transformed into an equation in the dimensionless variables
$kr$ such that $k$ does not appear anymore. This leads to the simple
recursion~(\ref{eq1.15}) for the coefficients. Here, in the matrix
case, the $k_n$ cannot easily be absorbed and the recursion formulae
for the coefficients will depend on the $k_n$ in a non--trivial way.
However, in some simple cases, it is possible to solve them. These
difficulties were an important motivation for us to develop the
methods which we introduced in Sec.~\ref{sec4}.

\section{Applications}
\label{sec8}

Can we obtain explicit formula for the radial functions
by using the recursion formula~(\ref{eq7.5}) ? --- At least in some
cases, this ought to be possible. Here, we present our first attempts. 

For the sake of completeness, we comment once more on the special case
$\beta=2$, i.e.~the unitary case. Obviously, the measure~(\ref{eq7.6})
simplifies enormously. This is so because the radial Gelfand--Tzetlin
coordinates coincide with the angular ones. Thus, the case $\beta=2$
is identical to the re--derivation of the Itzykson--Zuber integral by
Shatashvili~\cite{SLS}.

We now consider the orthogonal case $\beta =1$. The recursion formula
reads
\begin{eqnarray}
\Phi_N^{(1)}(x,k) &=&  G_N^{(1)}\int \frac{\Delta_{N-1}(x^\prime)}
                             {\sqrt{-\prod_{n,m}(x_n-x_m^\prime)}}  \cr
                  & &  \qquad \exp\left(i\left(\sum_{n=1}^Nx_n - 
                       \sum_{n=1}^{N-1}x^\prime\right)k_N\right) \, 
                       \Phi_{N-1}^{(1)}(x^\prime,\widetilde{k})
                       d[x^\prime] \ .
\label{usp0}
\end{eqnarray}
The square roots appearing in the measure make a further evaluation
very difficult. As obvious from the trivial case $N=2$, given in
Eq.~(\ref{eq1.31}), the function $\Phi_N^{(1)}(x,k)$ will be an {\it
  infinite} series for all values of $N$.  However, we expect that,
due to the different construction, this series is different from the
expansion in {\it zonal functions} which was obtained by
Muirhead~\cite{MUI}.

Obviously, there is a pattern emerging. The integration
measure~(\ref{eq7.6}) is purely {\it rational} for all even and
positive values of $\beta$. This is reminiscent of the situation for
vector Bessel functions in {\it odd} dimensions $d$, which consist of
a {\it finite} number of terms, as discussed in Sec.~\ref{sec1}.
Hence, we conjecture that the radial functions $\Phi_N^{(\beta)}(x,k)$
can also be written as a {\it finite} sum, exclusively containing
exponential and rational functions.

For all other values of $\beta$, the measure~(\ref{eq7.6}) is
algebraic, but not rational, and the radial functions
$\Phi_N^{(\beta)}(x,k)$ must be {\it infinite} series.  Nevertheless,
these infinite series contain exponential and rational functions.
Thus, they are different from expansions in terms of zonal
polynomials.

To furnish our conjecture about the form of the radial functions
for even and positive values of $\beta$ with an illustrative
example, we turn to the unitary--symplectic case $\beta=4$.
To simplify the notation we avoid the imaginary unit by writing
\begin{equation}
\Phi^{(4)}_{N}(-ix,k) \ = \ \int_{U\in {\it USp}(2N)}  
                      \exp\left(\Tr u^{-1}xuk\right) d\mu(U)\qquad,
\label{usp1}
\end{equation}
where $x=\diag(x_1,x_1,\dots,x_N,x_N)$ and
$k=\diag(k_1,k_1,\dots,k_N,k_N)$ are diagonal matrices with Kramers
degeneracies. The starting point of the recursion is the smallest
non--trivial case $N=2$, i.e.~the group $USp(4)$. We obtain after an
elementary calculation
\begin{eqnarray}
\lefteqn{\Phi^{(4)}_{2}(-ix,k) \ =}\cr
      &&      G_2^{(4)}\ \sum_{\omega\in S_2}
              \left(
              \frac{1}{\Delta_2^2\left(x\right)
                       \Delta_2^2\left(\omega(k)\right)}-
              \frac{2}{\Delta_2^3\left(x\right)
                       \Delta_2^3\left(\omega(k)\right)} 
              \right)
              \exp(\Tr x\omega(k))\quad .
\label{usp3}
\end{eqnarray}
The sum runs over the elements of the permutation group $S_N$ for
$N=2$.  Inserting Eq.~(\ref{usp3}) into the recursion formula, we find
for ${\it USp}(6)$, the next step in the recursion,
\begin{eqnarray}
\lefteqn{\Phi_{3}^{(4)}(-ix,k) \ = \ G_3^{(4)}G_2^{(4)}
                    \sum_{\omega\in S_2} \int_{x_1}^{x_2}
                    dx_1^\prime\int_{x_2}^{x_3}dx_2^\prime
                    \frac{\prod_{i=1}^3\prod_{j=1}^2 (x_i-x_j^\prime)}
                   {\Delta_3^3(x)\Delta_2^2(\omega(\widetilde{k}))}}\cr
             & & \exp\left((\Tr x-\Tr x^\prime)k_3+
                 \Tr x\omega(\widetilde{k})\right)
                 \left(\frac{1}{\Delta_2(x^\prime)}-
                 \frac{2}{\Delta_2^2(x^\prime)
                 \Delta_2(\omega(\widetilde{k}))}
                 \right)\quad .\cr
              &&
\label{usp4}
\end{eqnarray}
Although the integrand is finite everywhere, in particular at
$x_1^\prime=x_2^\prime=x_2$, the denominators $\Delta_2(x^\prime)$ and
$\Delta_2^2(x^\prime)$ raise a technical difficulty. The key to remove
them is to use the identity
\begin{equation}
\frac{2}{\Delta_2^2(x^\prime)} \ = \ -\left(
                              \frac{\partial}{\partial x_1^\prime}-
                              \frac{\partial}{\partial x_2^\prime}
                              \right)\frac{1}{\Delta_2(x^\prime)}
\label{usp5}
\end{equation}
and to observe that the product
$\prod_{i=1}^3\prod_{j=1}^2(x_i-x_j^\prime)$ annihilates all boundary
terms. Hence, we can integrate by parts and arrive at
\begin{eqnarray}
\Phi^{(4)}_{3}(-ix,k)&=&G_3^{(4)}G_2^{(4)}
                 \sum_{\omega\in S_2}
                 \frac{1}{\Delta_3^3(x)\Delta_2^3(\omega(\tilde{k}))}\cr
             & & \qquad\qquad\int_{x_1}^{x_2} dx_1^\prime
                             \int_{x_2}^{x_3} dx_2^\prime
                 \sum_{i=1}^3\prod_{j=1\atop j\neq i}^3 
                 (x_j-x_1^\prime)(x_j-x_2^\prime)\cr
             &&  \qquad\exp\left((\Tr x-\Tr x^\prime)k_3
                 +\Tr x\omega(\tilde{k})\right)\ ,
\label{usp6}
\end{eqnarray} 
where no denominator is left. Due to the permutation symmetry of the
original integral, we can restrict ourselves to the unity element {\it
  e} of the permutation group in the further evaluation of
Eq.~(\ref{usp6}). Thus we need only to consider the limits
$x_i^\prime\to x_i$, $i=1,2$ while integrating by parts.  After
collecting orders in $k$ we find
\begin{eqnarray}
\lefteqn{\Phi^{(4)}_{3,{\it e}}(-ix,k) \ = \ G_3^{(4)}G_2^{(4)}\frac{1}
                       {\Delta_3^3(x)\Delta_3^3(k)}}\cr
                   & & \left(-\Delta_3(x)\Delta_3(k)
                         +2\sum_{i<j}^3\left(
                         \frac{\Delta_3(x)\Delta_3(k)}
                         {(x_i-x_j)(k_i-k_j)}\right)-\right.\cr
                     & & \qquad\qquad\qquad\left.
                        4\sum_{i<j}^3(x_i-x_j)(k_i-k_j)+12\right)
                        \exp\left(\Tr x k\right) \ .
\label{usp8}
\end{eqnarray}
By introducing the composite variables
\begin{equation}
z_{\omega(ij)} \ = \ (x_i-x_j)(k_{\omega(i)}-k_{\omega(j)}) 
               \quad i,j = 1,\ldots,3,\quad\omega\in S_3,
\label{usp9}
\end{equation}
we can express $\Phi^{(4)}_{3}(-ix,k)$ in a compact form as
\begin{eqnarray}
\Phi^{(4)}_{3}(-ix,k)&=& G_3^{(4)}G_2^{(4)}\sum_{\omega\in S_3}
                    \frac{1}{\Delta_3^3(x)\Delta_3^3(\omega(k))}\cr
   &&\qquad\qquad\qquad  \left(4+\prod_{i<j}^3(2-z_{\omega(ij)})\right)
                    \exp\left(\Tr x \omega(k)\right)\ .
\label{usp10}
\end{eqnarray}
So far, we have not been able to extend this procedure to all
values of $N$.

However, we succeeded in calculating $\Phi_4^{(4)}(-ix,k)$, i.e.~the
case of the group $USp(8)$, by an hybrid method which combines
informations obtained from the recursion with an Hankel ansatz as
described in the previous section. We extend the right hand side of
Eq.~(\ref{usp10}) for $N=3$ to $N=4$ and use this expression as an
ansatz for the function $W_{N,\omega}^{(4)}(x,k)$. As it turns out, a
correction term is needed and, furthermore a correction to the
correction. Fortunately, there is a structure to this.  We give the
details in App.~\ref{appG}. We emphasize that the knowledge of
$\Phi^{(4)}_{3}(-ix,k)$ is essential for this hybrid procedure, in
particular the fact, that $\Phi^{(4)}_{3}(-ix,k)$ contains only linear
terms in every composite variable $z_{\omega(ij)}$.  Up to a
normalization, $\Phi_4^{(4)}(-ix,k)$ is given by
\begin{eqnarray}
\lefteqn{\Phi_4^{(4)}(-ix,k) \ = \ \sum_{\omega\in S_4}
              \frac{1}{\Delta_4^3(x)\Delta_4^3(\omega(k))}\Bigg(
              \prod_{i<j}(2- z_{\omega(ij)}) + \Big.}\cr
           &  &\qquad\qquad\qquad\qquad\qquad
              \frac{1}{2}\sum_{l<m<n}
              \prod_{i<j \atop {\neq lm \atop {\neq ln \atop \neq mn}}}
              (2- z_{\omega(ij)}) + \cr
           & &\Bigg.\qquad\qquad\qquad\qquad\frac{1}{4}
              \sum_{l<m \atop k<n}
          \prod_{i<j , \neq lk , \neq ln \atop {\neq mk ,\neq mn
          , \neq kn}}(2-z_{\omega(ij)})\Bigg)
          \exp\left(\Tr x\omega(k)\right)\ .  
\label{usp11}
\end{eqnarray}
Comparing this result with Eq.~(\ref{usp10}) we notice that, once
more, the composite variables $z_{\omega(ij)}$ enter only linearly in
the polynomial part of $\Phi_4^{(4)}(-ix,k)$.  Similarly, the
spherical Bessel function $j_1(z)$, which is the counterpart of
$\Phi_N^{(4)}(x,k)$ in the vector case given in Eqs.~(\ref{eq1.6}) and
(\ref{eq1.31}), has a polynomial part linear in $z$.  We expect that
such analogies are also present for higher values of $\beta$ and the
dimension $d$.

Formula~(\ref{usp11}) indicates a general structure for
$\Phi_N^{(4)}(x,k)$. The leading term is always the generating
function of the elementary symmetric functions in $z$. To this term
combinations of other symmetric functions are added, where certain
combinations of indices are cut out.

\section{Connection to Calogero--Sutherland Models}
\label{sec9}

The radial functions are related to, but different from, the
eigenfunctions which are usually employed in models of the
Calogero--Sutherland type. Since this issue is so important for
applications and so often raised in discussions, we briefly collect
the main points.

The radial Laplace operator defined in Eq.~(\ref{eq7.2}) is closely
related to the Calogero--Sutherland Hamiltonians.  In general one can
always cast a Fokker--Planck operator in a Hamilton operator by
adjunction~\cite{RIS} with the square root of the stationary
probability distribution defined through $\Delta_x P_{\rm eq}(x)=0$.
Choosing $P_{\rm eq}(x) = |\Delta_N(x)|^{-\beta/2}$,
the operator~(\ref{eq7.2}) can be associated with the Hamiltonian
\begin{equation}
H_D \ = \ -\sum_{n=1}^N\frac{\partial^2}{\partial x_n^2} \, + \,
       \frac{\beta(\beta-2)}{4}\sum_{n<m}^N\frac{1}{(x_n-x_m)^2} \ .
\label{cs3}
\end{equation}
It describes a scattering system with a continuous spectrum, the large
time behavior is determined by the states near the ground state.
Apart from a sign, this operator $H_D$ coincides with the operator
$L_{x,0}$ in Eq.~(\ref{eq1.34}) for $k=0$.  We also notice that 
the interaction vanishes for $\beta=2$.

To have a well defined thermodynamic limit one often confines the
motion of the particles to a circle.  This yields the
Calogero--Sutherland Hamiltonian
\begin{equation}
H_{CS} \ = \ -\sum_{n=1}^N\frac{\partial^2}{\partial x_n^2} \, + \,
       \frac{\beta(\beta-2)}{4}\sum_{n<m}^N\frac{(\pi/N)^2}
        {\sin^2\left(\pi(x_n-x_m)/N\right)}\quad .
\label{cs4}
\end{equation}
which can also be derived directly from Dyson's circular ensembles
\cite{Mehta}. 
Another way of confining the particles is by a harmonic
potential. This leads to the Calogero Hamiltonian~\cite{CAL} 
\begin{equation}
H_{C} \ = \ -\sum_{n=1}^N\frac{\partial^2}{\partial x_n^2} \, + \,
       \frac{\beta(\beta-2)}{4}\sum_{n<m}^N\frac{1}{(x_m-x_m)^2} \, + \,
       \frac{1}{16}\sum_{n=1}^N x_n^2 \ .
\label{cs2}
\end{equation}
In the thermodynamic limit the particle density $R_1(x)$ of the ground
state is described by Wigner's semi-circle law.  The mean particle
level spacing $D=1/R_1(0)$ scales as $D \propto 1/\sqrt{N}$. Therefore
in the thermodynamic limit the harmonic confining term in
Eq.~(\ref{cs2}) vanishes on the scale of the mean level spacing. On
this {\it unfolded scale} the correlation functions become independent
of the confinement mechanism.  The three Hamiltonians $H_C, H_D$ and
$H_{CS}$ are known to be integrable systems for arbitrary $\beta$
\cite{SU1}.  However, the three values $\beta=1,2,4$ are
distinguished, since they establish a connection to the random matrix
ensembles. Indeed, for these values of $\beta$ they belong to a much
wider class of integrable systems, which can be constructed by means
of the root space of a simple Lie algebra or -- still more generally
-- of a Kac--Moody algebra~\cite{OP}.  This class comprises
Hamiltonians which can be derived by an adjunction procedure from a
Laplace--Beltrami operator of a group acting in a symmetric space.
This space has positive curvature for $H_{CS}$ and zero curvature for
$H_D, H_{C}$.  In Refs.~\cite{BEE,CAS} it was pointed out, that the
Dorokhov--Mello--Pereyra--Kumar equation for scattering matrices with
broken time reversal symmetry corresponds to a Laplace--Beltrami
operator in a symmetric space of negative curvature.

Eigenfunctions $\psi^{(\beta)}_{N,E}(x)$ of the Hamiltonians $H_C,
H_D, H_{CS}$ with eigenenergy $E$ for arbitrary $\beta$ are known.
Essentially, these solutions are products of the ground state wave
function and symmetric polynomials in the coordinates $x$ of the $N$
particles. In case of the Calogero--Sutherland Hamiltonian $H_{CS}$,
these polynomials are the Jack polynomials~\cite{STA,FOR,HA}. In this
approach, the energy eigenvalues $E$ are labeled by a partition of
length $N$.  The crucial difference to the matrix Bessel functions is
that the Jack polynomials are symmetric polynomials in one set of
variables $x$ only whereas the matrix Bessel functions
$\Phi_N^{(\beta)}(x,k)$ are symmetric in two sets of variables $x$ and
$k$.  Importantly, they are, in addition, symmetric under interchange
of the two sets of variables,
$\Phi_N^{(\beta)}(x,k)=\Phi_N^{(\beta)}(k,x)$.  This is reflected in
the fact that the operator $L_{x,\omega(k)}$ emerging in the Hankel
ansatz depends on $k$ while $H_D$ does not.  Due to their symmetry,
the matrix Bessel functions $\Phi_N^{(\beta)}(x,k)$ are, at least for
$H_D$, the more natural eigenfunctions. This becomes obvious in the
fact that, to obtain orthogonality conditions, one has to sum the
$\psi^{(\beta)}_{N,E}(x)$ over an infinite number of partitions.  On
the other hand, orthogonality relations are an inherent feature of the
$\Phi_N^{(\beta)}(x,k)$ due to their meaning in the Fourier--Bessel
analysis, as discussed in App.~\ref{appFBA}.

In other words, the functions $\psi^{(\beta)}_{N,E}(x)$ can be viewed
as a basis in an expansion of the $\Phi_N^{(\beta)}(x,k)$.  We can
consider the variables $k$ as a set of real numbers corresponding to
the energies $E$ labeling the eigenstates of $H_D$.  The matrix Bessel
functions~(\ref{eq1.27a}) are solutions of the Schr\"odinger equation
with Hamiltonian $H_D$ for the coupling parameters $\beta=1,2,4$. The
recursion formula~(\ref{eq7.5}) represents an analytic continuation of
these integral solutions to arbitrary positive $\beta$. All these
functions $\Phi_N^{(\beta)}(x,k)$ have, for arbitrary $\beta$
additional features, such as the symmetry in $x$ and $k$, which have
no analogue in the functions $\psi^{(\beta)}_{N,E}(x)$. The merit of
our recursion formula lies in the fact that, a priori, no infinite
resummation is required to obtain functions of the type
$\Phi_N^{(\beta)}(x,k)$. Nevertheless Forrester~\cite{FOR} and Nagao
and Forrester~\cite{NF} showed that such resummed expressions can
successfully be used in certain cases. They treated the case of
Poissonian initial conditions~\cite{NF} for the Calogero--Sutherland
Hamiltonian $H_{CS}$ and derived exact expressions for the correlation
functions for arbitrary $\beta$ for one or two particles. This is also
related to the works of Muirhead~\cite{MUI} and Pandey~\cite{PAN}.

\section{Summary and Conclusion}
\label{sec10}

We presented a recursive construction for certain spherical functions.
We referred to them as matrix Bessel functions because, first, they
are a natural extension of vector Bessel functions in the sense that
the integration over a group corresponds to the integration over a
solid angle and, second, they satisfy a partial differential equation
generalizing the Bessel ordinary differential equation. For matrices,
the index $\beta$ labeling the groups appears analogously to the
dimension $d$ in the case of vectors. The introduction of radial
Gelfand--Tzetlin coordinates, which are related to but different from
the ordinary angular ones, was crucial for the recursion.  The Cayley
transformation ought to provide a connection between the angular and
the radial Gelfand--Tzetlin coordinates~\cite{GO}.  As evident from
its construction, the recursion maps an integral over a group fully
onto an iteration which exclusively takes place in the radial space.

Remarkably, the recursion turned out to be far more general than was
to be expected, at first sight, from the proof which involved Lie
groups. We showed that our recursion is also the iterative solution of
the corresponding partial differential equation for arbitrary values
of $\beta$. We introduced the term radial functions for this
generalization of matrix Bessel functions.  We expect that one has to
employ the theory of quantum groups to give a group theoretical
derivation of the recursion formula for arbitrary values of $\beta$.

Using the recursion formula, we discussed the structure of radial
Bessel functions. We conjectured that, for even $\beta$, they can be
written as finite sums involving only exponential and rational
functions. We illustrated that by working out, for $\beta=4$, the
cases of $N=3$ and $N=4$ distinct eigenvalues. Further evaluation of
explicit formulae for arbitrary $N$ and, maybe, for all even $\beta$
does not seem impossible. Work is in progress.  The extension of the
stationary phase approach by Duistermaat and Heckman~\cite{DH} to
higher orders could, for even $\beta$, be an alternative to derive
such explicit results, because the expansion terminates.  In this
context, we mention that the radial function for higher values of
$\beta$ are, to some extent, but not fully, the higher order radial
functions for lower values of $\beta$. This also generalizes the
situation for ordinary Bessel functions. However, there are many more
higher order radial functions, they are not at all exhausted by this
mapping between values of $\beta$.

In the present contribution, we only focussed on ordinary spaces,
i.e.~spaces which are built upon commuting numbers. In a second
study~\cite{GuKo} we also address superspaces which involve commuting
and anticommuting variables.

\section*{acknowledgement}

We thank Y.~Fjodorov, F.~Leyvraz, G.~Olshanski, W.~Schempp,
T.~Seligman and R.~Weissauer for fruitful discussions at various
stages of this work and for pointing out important references to us.
We acknowledge financial support from the Deutsche
Forschungsgemeinschaft, TG as a Heisenberg fellow and HK as a doctoral
stipend, HK also thanks the Max--Planck-Institute for financial
support.

\appendix

\section{Alternative Integral Representation for 
                     Vector Bessel Functions}
\label{appVBF}
\renewcommand{\theequation}{A.\arabic{equation}}
\setcounter{equation}{0}

The Bessel functions $\chi^{(d)}(kr)$ are defined as an integral over
angles in Eq.~(\ref{eq1.1}), but they can also be written as integrals
over the entire or half real axis~\cite{AS}.  To make possible an
instructive comparison with the matrix case, we quote and re--derive
the representation
\begin{equation}
\chi^{(d)}(kr) \ = \ \frac{\Gamma(d/2)}{i2\pi} 
       \left(\frac{2}{ikr}\right)^{(d-2)/2} \int_{-\infty}^{+\infty} 
       \exp\left(-i\frac{kr}{2}\left(t+\frac{1}{t}\right)\right)
       \frac{dt}{t^{d/2}} \ .
\label{eq1.16}
\end{equation}
The singularities have to be treated properly. 

The position vector in the $d$ dimensional space is
$\vec{r}=r\vec{e}_r$ where the vector $\vec{e}_r$ parametrizes the
unit sphere.  To integrate over its orientation, i.e.~over the solid
angle $\Omega$, one can re--express the measure as
\begin{equation}
d\Omega \ = \ \frac{\Gamma(d/2)}{\pi^{d/2}} \, 
              \delta\left(\vec{e}_r^2-1\right) \, d^de_r \ .
\label{eqVBF.1}
\end{equation}
Here, the vector $\vec{e}_r$ is re--interpreted: its components live
on the entire real axis and the domain of integration is the full $d$
dimensional space with the Cartesean measure $d^de_r$. The $\delta$
distribution confines the vector to the unit sphere.  Writing this
distribution as a Fourier transform, we obtain from Eq.~(\ref{eq1.1})
\begin{eqnarray}
\chi^{(d)}(kr) &=& \frac{\Gamma(d/2)}{\pi^{d/2}} \, \frac{1}{2\pi}
                     \int_{-\infty}^{+\infty} dt \int d^de_r \,
                     \exp\left(it\left(\vec{e}_r^2-1\right) \right) \, 
                     \exp\left(i\vec{k}\cdot r\vec{e}_r\right) 
                                     \nonumber\\ 
               &=& \frac{\Gamma(d/2)}{2\pi}
      \int_{-\infty}^{+\infty} \frac{\exp(-it)}{(it)^{d/2}}
            \exp\left(-i\frac{(kr)^2}{4t}\right) dt
\label{eqVBF.2}
\end{eqnarray}
where the integral over $\vec{e}_r$ gave a Gaussian in $d$ dimensions.
The contour for the integration over $t$ has to be chosen
appropriately. Upon a trivial change of variables, this result
yields Eq.~(\ref{eq1.16})

\section{Alternative Integral Representation for 
                     Matrix Bessel Functions}
\label{appMBF}
\renewcommand{\theequation}{B.\arabic{equation}}
\setcounter{equation}{0}

The matrix Bessel functions $\Phi_N^{(\beta)}(x,k)$ for $\beta=1,2,4$
can be be written in an alternative way. Although we can hardly
believe that this representation is completely new, we could not find
it in the literature. Similarly to Eq.~(\ref{eq1.16}) in the vector
case, we can write
\begin{equation}
\Phi_N^{(\beta)}(x,k) \ = \ A_N^{(\beta)} \int d[T] 
      \exp\left(i\Tr T\right) \,
      \Det^{-\beta/2}\left(x\otimes\hat{k} - T\otimes 1_N\right)
\label{eq1.38}
\end{equation}
where $1_N$ is the $N\times N$ unit matrix. The normalization constant
is given by
\begin{equation}
A_N^{(\beta)} \ = \ \frac{i^{\beta N^2/2} \pi^{\beta N(N-1)/4}}
                         {\beta^{N+\beta N(N-1)/2}} \,
                         \prod_{n=1}^N \Gamma(\beta n/2) \ .
\label{eq1.38a}
\end{equation}
The matrix $T$ in Eq.~(\ref{eq1.38}) is real symmetric, Hermitean or
Hermitean self--dual, respectively, for $\beta=1,2,4$.  The measure
$d[T]$ is Cartesean and given by Eq.~(\ref{eq1.22a}). All independent
variables in $T$ are integrated over the entire real axis.  To ensure
convergence, the diagonal elements of $T$ have to be given a proper
imaginary increment.  We notice that $x$ and $T$ are $N\times N$
matrices for $\beta=1,2$ and $2N\times 2N$ for $\beta=4$ with doubly
degenerated eigenvalues.  The matrix $\hat{k}$ is, in all three cases
$\beta$, just the $N \times N$ matrix
$\hat{k}=\diag(k_1,k_2,\ldots,k_N)$, as defined following
Eq.~(\ref{eq1.24}).

The integral representation~(\ref{eq1.38}) leads to an interesting
integral equation for the matrix Bessel functions,
\begin{equation}
\Phi_N^{(\beta)}(x,k) \ = \ B_N^{(\beta)} \, \Det^{1-\beta/2}x \,
      \int d[t] |\Delta_N(t)|^\beta 
      \frac{\Phi_N^{(\beta)}(x,t)} 
           {\prod_{n,m}\left(k_m-t_n\right)^{\beta/2}} \ ,
\label{eq1.39}
\end{equation}
where the normalization constant reads
\begin{equation}
B_N^{(\beta)} \ = \ \frac{i^{\beta N^2/2} \Gamma^N(\beta/2)}
                         {(2\pi)^N N!} \ .
\label{eq1.39a}
\end{equation}
The $t_n$ in Eq.~(\ref{eq1.39}) have a proper imaginary increment and
their domain of integration is the real axis. Due to the symmetry
relation~(\ref{eq1.27b}), the variables $x$ and $k$ can be
interchanged in Eqs.~(\ref{eq1.38}) and~(\ref{eq1.39}). 

It is not difficult to see from the integral equation~(\ref{eq1.39})
that the product in the denominator of its right hand side can, in
the case $\beta=2$ be written as
\begin{equation}
\frac{B_N^{(2)}}{\prod_{n,m}\left(k_m-t_n\right)}
\ = \ \frac{\det[\delta(x_n-t_m)]_{n,m=1,\ldots,N}}
           {|\Delta_N(k)\Delta_N(t)|^{1/2}} \ . 
\label{eq1.40}
\end{equation}
For $\beta\neq 2$, the term $\Det^{1-\beta/2}x$ contributes.
Nevertheless, the product still shares features with a $\delta$
distribution.

To derive this alternative integral representation, we proceed
analogously to Eq.~(\ref{eqVBF.1}) by re--writing the invariant
measure of $U\in U(N;\beta)$ using $\delta$ distributions. The
invariance simply means that all columns $U_n, \ n=1,\ldots,N$ are
orthonormal, $\Tr U_n^\dagger U_m=\delta_{nm}$. The trace $\Tr$ is
only needed for $\beta=4$, because the entries of $U$ are quaternions
in this case. Thus, we may write
\begin{equation}
d\mu(U) \ = \ M_N^{(\beta)} \, d[U]
   \prod_{n=1}^N \delta\left(\Tr U_n^\dagger U_n-1\right)
   \prod_{n<m} \delta\left(\Tr U_n^\dagger U_m\right)
\label{eqMBF.1}
\end{equation}
where $d[U]$ is the Cartesean measure of all entries of $U$ and the
integration is for all variables over the entire real axis.  The
constant $M_N^{(\beta)}$ will be determined later.
Ullah~\cite{UL1,UL2} used such forms for the measure to work out
certain probability density functions.  The bilinear forms in the
$\delta$ distributions have $\beta$ components for $n\ne m$,
\begin{equation}
U_n^\dagger U_m \ = \ \sum_{\alpha=0}^{\beta-1}
                      \left[U_n^\dagger U_m\right]^{(\alpha)} 
                      \tau^{(\alpha)} \ .
\label{eqMBF.2}
\end{equation}
We notice that $\left[U_n^\dagger U_n\right]^{(\alpha)}=0$ for
$\alpha>0$ in the case $n=m$, because the length of every vector is
real. Thus, because of Eq.~(\ref{eqMBF.2}), the $\delta$ distributions
in the measure~(\ref{eqMBF.1}) have to be products of $\delta$
distributions for every non--zero component $\left[U_n^\dagger
U_m\right]^{(\alpha)}$. We now introduce Fourier representations
\begin{eqnarray}
\delta\left(\left[U_n^\dagger U_m\right]^{(\alpha)}\right)
             &=& \frac{1}{\pi} \int_{-\infty}^{+\infty}
                 dT_{nm}^{(\alpha)} 
   \exp\left(-i2\left[U_n^\dagger U_m\right]^{(\alpha)}
                                    T_{nm}^{(\alpha)}\right) 
                                          \nonumber\\
\delta\left(\left[U_n^\dagger U_n\right]^{(0)}-1\right)
             &=& \frac{1}{2\pi} \int_{-\infty}^{+\infty}
                 dT_{nn}^{(0)} 
   \exp\left(-i\left(\left[U_n^\dagger U_n\right]^{(0)}-1\right)
                                         T_{nm}^{(0)}\right) 
\label{eqMBF.3}
\end{eqnarray}
for $n\ne m$ and $n=m$, respectively. The Fourier variables form the
elements
\begin{equation}
T_{nm} \ = \ \sum_{\alpha=0}^{\beta-1} T_{nm}^{(\alpha)} \, 
                                 \tau^{(\alpha)} \ .
\label{eqMBF.4}
\end{equation}
of a matrix $T$ which is real--symmetric, Hermitean or Hermitean
self--dual according to $\beta=1,2,4$. We notice that the diagonal
elements $T_{nn}=T_{nn}^{(0)}$ are always real.
\begin{eqnarray}
\delta\left(\Tr U_n^\dagger U_m\right) 
             &=& \frac{1}{\pi^\beta} \int
                 d^\beta T_{nm}
   \exp\left(-i\Tr U_n^\dagger(T_{nm}\otimes 1_N) U_m \right.
                                          \nonumber\\
             & & \qquad\qquad\qquad\qquad
       \left.-i\Tr U_m^\dagger(T_{nm}^*\otimes 1_N) U_n\right)
                                          \nonumber\\
\delta\left(\Tr U_n^\dagger U_n-1\right)
             &=& \frac{1}{2\pi} 
                 \int_{-\infty}^{+\infty} dT_{nn}
   \exp\left(i\Tr T_{nn}
             -i\Tr U_n^\dagger(T_{nn}\otimes 1_N) U_n\right)
\label{eqMBF.5}
\end{eqnarray}
for $n\ne m$ and $n=m$, as above. Just as the trace $\Tr$, the direct
product is only needed in the case $\beta=4$.

We order the columns $U_n, \ n=1,\ldots,N$ of the matrix $U$ in a
vector $\vec{U}=(U_1,U_2,\ldots,U_N)^T$ with $N^2$ elements. For
$\beta=1,2$, the elements are scalars, for $\beta=4$, they are
quaternions. Collecting everything, we can re--write the
measure~(\ref{eqMBF.1}) in the form
\begin{equation}
d\mu(U) \ = \ \frac{M_N^{(\beta)} d[U]}
                   {(2\pi)^N\pi^{\beta N(N-1)/2}} 
              \int d[T] \exp\left(i\Tr T 
           -i\Tr\vec{U}^\dagger (T\otimes 1_N)\vec{U}\right) \ .
\label{eqMBF.6}
\end{equation}
To use this in the integral~(\ref{eq1.27a}) for the matrix Bessel
functions $\Phi_N^{(\beta)}(x,k)$, we also take advantage of the
relation
\begin{equation}
\Tr U^{-1}xUk \ = \ \Tr \vec{U}^\dagger (x\otimes\hat{k})\vec{U}  
\label{eqMBF.7}
\end{equation}
which allows us to write
\begin{eqnarray}
\Phi_N^{(\beta)}(x,k) &=& \frac{M_N^{(\beta)}}
                   {(2\pi)^N\pi^{\beta N(N-1)/2}} 
               \int d[U] \int d[T] \exp\left(i\Tr T\right)
                               \nonumber\\ 
           & & \qquad\qquad \qquad\qquad 
  \exp\left(i\Tr\vec{U}^\dagger(x\otimes\hat{k} - 
                            T\otimes 1_N)\vec{U}\right) 
                               \nonumber\\ 
           &=& \frac{M_N^{(\beta)}i^{\beta N^2}
                     \pi^{\beta N/2}}{(2\pi)^N} 
               \int d[T] \exp\left(i\Tr T\right)
                               \nonumber\\ 
           & & \qquad\qquad \qquad\qquad 
      \Det^{-\beta/2}\left(x\otimes\hat{k} - T\otimes 1_N\right) \ .
\label{eqMBF.8}
\end{eqnarray}
Thus, the integration over $U$ could be done as a Gaussian one and
gave the result~(\ref{eq1.38}). Obviously, the Gaussian integrals over
$\vec{U}$ only converge, if the diagonal elements of $T$ have a proper
imaginary increment.

Formula~(\ref{eqMBF.8}) yields immediately the integral
equation~(\ref{eq1.39}). Upon making the change of variables
\begin{equation}
T \ = \ x^{1/2}T^\prime x^{1/2} \ , \qquad {\rm implying} \qquad
d[T] \ = \ \Det^{1+\beta(N-1)/2}x \, d[T^\prime] \ ,
\label{eqMBF.8a}
\end{equation}
we bring $x$ into the exponential function and remove it from
the determinant. We diagonalize 
$T^\prime=V^{\prime -1}t^\prime V^\prime$ and find
\begin{equation}
\Det^{-\beta/2}\left(x\otimes\hat{k} - T\otimes 1_N\right) \ = \
   \Det^{-\beta N/2}x \, \prod_{n,m}(k_m-t_n^\prime)^{-\beta/2} \ .
\label{eqMBF.8b}
\end{equation}
The integral over $V^\prime$ is then just the integral
definition~(\ref{eq1.27a}) of the matrix Bessel function
$\Phi_N^{(\beta)}(x,t^\prime)$ and we arrive at Eq.~(\ref{eq1.39}).

The normalization constants remain to be derived. Conveniently,
they nicely relate to a special form of Selberg's integral which is
given in Eq.~(17.5.2) of Mehta's book~\cite{Mehta},
\begin{eqnarray}
J_N &=& \int d[t] |\Delta_N(t)|^{2\gamma}
                  \prod_{n=1}^N\left(a_1+it_n\right)^{-b_1}
                               \left(a_2-it_n\right)^{-b_2}
                                          \nonumber\\
    &=& \frac{(2\pi)^N}{(a_1+a_2)^{(b_1+b_2)N-\gamma N(N-1)-N}} \,
                                          \nonumber\\
    & & \qquad\qquad
        \prod_{n=0}^{N-1} \frac{\Gamma(1+(n+1)\gamma)
                                \Gamma(b_1+b_2 -(N+n-1)\gamma -1)}
           {\Gamma(1+\gamma)\Gamma(b_1-n\gamma)\Gamma(b_2-n\gamma)} \ .
\label{eqMBF.9}
\end{eqnarray}
We now put $x=0$ or $k=0$ and have $\Phi_N^{(\beta)}(0,k)=1$ or
$\Phi_N^{(\beta)}(x,0)=1$ on the left hand side of
Eq.~(\ref{eqMBF.8}). We diagonalize $T=V^{-1}tV$ and use the
invariance of the integral. Employing the measure~(\ref{eq1.23a}) and
the constant $C_N^{(\beta)}$ given in Eq.~(\ref{eq1.23c}), we find the
condition
\begin{equation}
1 \ = \ \frac{M_N^{(\beta)}C_N^{(\beta)}\pi^{\beta N/2}}
                                              {(2\pi)^N}
         \int d[t] |\Delta_N(t)|^\beta 
     \prod_{n=1}^N \frac{\exp(it_n)}{(it_n)^{\beta N/2}} \ .
\label{eqMBF.10}
\end{equation}
We map this onto Selberg's integral~(\ref{eqMBF.9}) by setting
$\gamma=\beta/2$, $b_1=\beta N/2$ and $a_2=b_2$, by using
\begin{equation}
\lim_{a_2\to\infty} \frac{a_2^{a_2}}{(a_2-it_n)^{a_2}} 
                               \ = \ \exp(it_n)
\label{eqMBF.11}
\end{equation}
and by considering $a_2^{Na_2}J_N$ in the limits $a_1\to 0$ and
$a_2\to\infty$. With the help of some standard asymptotic formulae for
the $\Gamma$ function, we obtain $M_N^{(\beta)}$ and, eventually, the
constants $A_N^{(\beta)}$ and $B_N^{(\beta)}$ in Eqs.~(\ref{eq1.38a})
and~(\ref{eq1.39a}).

\section{Fourier--Bessel Analysis}
\label{appFBA}
\renewcommand{\theequation}{C.\arabic{equation}}
\setcounter{equation}{0}

The Fourier--Bessel Analysis involving matrix Bessel functions was
discussed by Harish--Chandra~\cite{HC} in a general and formal way.
To show the connection to our results, we summarize here some
essential features of the Fourier--Bessel analysis on an explicit
level.

We write the Fourier transform of a function $f(H)$ as
\begin{equation}
F(K) \ = \ D_N^{(\beta)} \int d[H] \exp\left(i\Tr HK\right) f(H)
\label{eq1.50}
\end{equation}
where the matrices $H$ and $K$ have the same symmetries. If we choose
a symmetric normalization,
\begin{equation}
D_N^{(\beta)} \ = \ \frac{1}{(2\pi)^{N/2}\pi^{\beta N(N-1)/4}} \ ,
\label{eq1.51}
\end{equation}
we can write the inverse transform as
\begin{equation}
f(H) \ = \ D_N^{(\beta)} \int d[K] \exp\left(-i\Tr KH\right) 
                                   F(K) \ .
\label{eq1.52}
\end{equation}
We notice that, according to Eq.~(\ref{eq1.24a}), the Fourier
transform of the constant $D_N^{(\beta)}$ is the $\delta$ distribution
$\delta(K)$ and vice versa. 

If $f$ is an invariant function such that $f(H)=f(x)$, its Fourier
transform turns out to be invariant as well, $F(K)=F(k)$. Introducing
eigenvalue--angle coordinates, we easily find
\begin{equation}
F(k) \ = \ D_N^{(\beta)} C_N^{(\beta)} 
           \int d[x] \, |\Delta_N(x)|^\beta \, 
                        \Phi_N^{(\beta)}(x,k) \, f(x) 
\label{eq1.53}
\end{equation}
for the Fourier transform and
\begin{equation}
f(x) \ = \ D_N^{(\beta)} C_N^{(\beta)} 
           \int d[k] \, |\Delta_N(k)|^\beta \, 
                        \Phi_N^{(\beta)*}(k,x) \, F(k) 
\label{eq1.54}
\end{equation}
for its inverse. We now insert the transform~(\ref{eq1.53}) into
the inverse~(\ref{eq1.54}) and conclude that
\begin{eqnarray}
& & \left(D_N^{(\beta)} C_N^{(\beta)}\right)^2
           \int d[k] \, |\Delta_N(k)|^\beta \, 
           \Phi_N^{(\beta)}(x,k) \, \Phi_N^{(\beta)*}(k,y) \nonumber\\
& & \qquad\qquad\qquad = \ \frac{\det[\delta(x_n-y_m)]_{n,m=1,\ldots,N}}
            {|\Delta_N(x)\Delta_N(y)|^{\beta/2}} \ .
\label{eq1.55}
\end{eqnarray}
This is the analogue of Hankel's expansion of the $\delta$
distribution. {}From Eq.~(\ref{eq1.55}), the formula
\begin{equation}
\int d\mu(U) \, \delta\left(U^\dagger xU - y\right)
\ = \ \frac{1}{C_N^{(\beta)}} \, 
              \frac{\det[\delta(x_n-y_m)]_{n,m=1,\ldots,N}}
                   {|\Delta_N(x)\Delta_N(y)|^{\beta/2}} \ .
\label{eq1.56}
\end{equation}
obtains. To see this, we introduce a matrix $G$ having the same
symmetries as $H$ and write
\begin{equation}
\delta(H-G) \ = \ (D_N^{(\beta)})^2 
                  \int d[K] \exp\left(-i\Tr K(H-G)\right) \ .
\label{eqFBA.1}
\end{equation}
Averaging over the diagonalizing matrix $U$ of $H$ yields
\begin{equation}
\int d\mu(U) \, \delta(H-G) \ = \ (D_N^{(\beta)})^2 
                  \int d[K] \, \Phi_N^{(\beta)*}(x,k)
                            \, \exp\left(i\Tr K G\right) \ ,
\label{eqFBA.2}
\end{equation}
by using the invariance of the measure. We now introduce
eigenvalue--angle coordinates for $K$ and do the integral over $V$,
the diagonalizing matrix of $K$,
\begin{equation}
\int d\mu(U) \, \delta(H-G) \ = \ (D_N^{(\beta)})^2 C_N^{(\beta)}
                  \int d[k] \, |\Delta_N(k)|^\beta \, \Phi_N^{(\beta)*}(x,k)
                            \, \Phi_N^{(\beta)}(k,y) \ ,
\label{eqFBA.3}
\end{equation}
where we have, once more, employed the invariance of the measure.
Since the right hand side of this equation does only depend on the
eigenvalues $y$ of $G$, we may replace $G$ on the left hand side with
$y$. Together with Eq.~(\ref{eq1.55}), this gives formula~(\ref{eq1.56}).

For the convolution in matrix space of two functions $f_1(H)$ and
$f_2(H)$, we straightforwardly find the generalization of the standard
convolution theorem,
\begin{equation}
f(H) \ = \ \int d[G] f_1(G) f_2(H-G)
     \ = \ \int d[K] \exp\left(-i\Tr HK\right) F_1(K) F_2(K)
\label{eq1.57}
\end{equation}
where $G$ has the same symmetries as $H$. The functions $F_1(K)$ and
$F_2(K)$ are the Fourier transforms of $f_1(H)$ and $f_2(H)$,
respectively. If the functions are invariant, the second of
Eqs.~(\ref{eq1.57}) acquires the form
\begin{equation}
f(x) \ = \ C_N^{(\beta)} \int d[k] \, |\Delta_N(k)|^\beta \,
                         \Phi_N^{(\beta)*}(x,k) \, F_1(k) \, F_2(k) \ .
\label{eq1.58}
\end{equation}
On the other hand, we find from the first of Eqs.~(\ref{eq1.57})
\begin{equation}
f(x) \ = \ C_N^{(\beta)} \int d[y] \, |\Delta_N(y)|^\beta \,
                         f_1(y) \, \hat{f}_2(x,y)
\label{eq1.59}
\end{equation}
where $y$ are the eigenvalues of $G$. This formula is a convolution in
the curved space of the eigenvalues. The second function is given by
\begin{equation}
\hat{f}_2(x,y) \ = \ \int d\mu(U) \, f_2(x-U^\dagger yU) \ .
\label{eq1.60}
\end{equation}
We insert the Fourier integral for $F_1(k)$ according to
Eq.~(\ref{eq1.53}) into Eq.~(\ref{eq1.58}), compare with
Eq.~(\ref{eq1.59}) and obtain the Fourier decomposition
\begin{equation}
\hat{f}_2(x,y) \ = \ D_N^{(\beta)}C_N^{(\beta)} 
                     \int d[k] \, |\Delta_N(k)|^\beta \,
                         \Phi_N^{(\beta)*}(x,k) \, F_2(k) \,
                         \Phi_N^{(\beta)}(k,y) \ .
\label{eq1.61}
\end{equation}
Formulae~(\ref{eq1.55}) and~(\ref{eq1.56}) can be viewed as special
cases of these results.

\section{Action of the Laplacean on the Radial Functions 
                     for Arbitrary $\beta$}
\label{appLIR}
\renewcommand{\theequation}{D.\arabic{equation}}
\setcounter{equation}{0} 

We make the notation more compact by defining
\begin{equation}
d\widetilde{\mu}(x^\prime,x)= d\mu(x^\prime,x)
                  \exp\left(i\left(\sum_{n=1}^Nx_n - 
                  \sum_{n=1}^{N-1}x_n^\prime\right)k_N\right)\quad ,
\label{Arad3}
\end{equation}
where the measure is given in Eq.~(\ref{eq7.6}).  To prove the
identity~(\ref{eq7.8}), we write the integral using $\Theta$
functions. The left hand side of Eq.~(\ref{eq7.8}) reads
\begin{equation}
\Delta_x\int \widetilde{\mu}(x^\prime,x) 
 \Phi_{N-1}^{(\beta)}(x^\prime,\widetilde{k})
 \prod_{i>j}\Theta(x_i-x_j^\prime)
 \prod_{j\ge l}\Theta(x_j^\prime-x_l) d[x^\prime]\ ,
\label{rad3}
\end{equation}
where now the integration domain is the real axis for all variables.
Thus, we can directly calculate the action of the operator $\Delta_x$
onto the integral. We find
\begin{eqnarray}
\lefteqn{\Delta_x\int \widetilde{\mu}(x^\prime,x) 
           \Phi_{N-1}^{(\beta)}(x^\prime,\widetilde{k})
           \prod_{i> j}\Theta(x_i-x_j^\prime)
           \prod_{j\ge l}\Theta(x_j^\prime-x_l)d[x^\prime]= }\cr
 &&        \qquad\qquad\qquad\qquad
           \int \Phi_{N-1}^{(\beta)}(x^\prime,\widetilde{k})
           \prod_{i> j}\Theta(x_i-x_j^\prime)
           \prod_{j\ge l}\Theta(x_j^\prime-x_l)\cr
 &&        \qquad\qquad\qquad\qquad
           \left(\Delta_{x^\prime}^{(-)}+\beta\sum_{n\neq m}
           \frac{1}{(x_n^\prime-x_m^\prime)^2}-k_N^2\right)
           \widetilde{\mu}(x^\prime,x)d[x^\prime]\cr
 &&        +\int \Phi_{N-1}^{(\beta)}(x^\prime,\widetilde{k})
           \widetilde{\mu}(x^\prime,x)
           \Delta_x\prod_{i>j}\Theta(x_i-x_j^\prime)
           \prod_{j\ge l}\Theta(x_j^\prime-x_)
           d[x^\prime]\cr
 &&        +2 \int \Phi(x^\prime,\widetilde{k})\sum_{n=1}^N
           \frac{\partial}{\partial x_n}\widetilde{\mu}(x^\prime,x)
           \frac{\partial}{\partial x_n}
           \prod_{i> j}\Theta(x_i-x_j^\prime)
           \prod_{j\ge l}\Theta(x_j^\prime-x_l)
           d[x^\prime]\ ,
\label{rad6}
\end{eqnarray}
where we define the operator
\begin{equation}
\Delta_{x^\prime}^{(-)}= \ \sum_{n=1}^N \frac{\partial^2}{\partial x_n^2}
               \, - \, \sum_{n<m} \frac{\beta}{x_n-x_m} 
               \left(\frac{\partial}{\partial x_n}-
                     \frac{\partial}{\partial x_m}\right) \ .
\label{rad4}
\end{equation}
By a series of integrations by parts, the operator
$\Delta_{x^\prime}^{(-)}$ acting on $\widetilde{\mu}(x^\prime,x)$ is
transformed to $\Delta_{x^\prime}$ acting only on
$\Phi(x^\prime,\widetilde{k})$.  At taking the derivative of the
$\Theta$ functions, we notice that only adjacent levels contribute,
because otherwise terms like $\Theta(x_i-x_j)$ with $i < j$ arise
which annihilate the integral due to the chosen ordering.  Therefore,
we can write
\begin{eqnarray}
\lefteqn{\frac{\partial}{\partial x_n}\prod_{i> j}\Theta(x_i-x_j^\prime)
         \prod_{j\ge l}\Theta(x_j^\prime-x_l)=}\cr
    &&\qquad\qquad\prod(\Theta_{\neq nn^\prime \ , \neq (n-1)^\prime n})
      \Big(\delta(x_n-x_n^\prime)\Theta(x_{n-1}^\prime-x_n) \Big. \cr
    &&\qquad\qquad\qquad\qquad\qquad
       \Big. - \delta(x_{n-1}^\prime-x_n)\Theta(x_n-x_n^\prime)\Big) \ ,
\label{rad7}
\end{eqnarray}
where $\prod(\Theta_{\neq nn^\prime} , \Theta_{\neq (n-1)^\prime n})$
is short--hand for the product on the left hand side of
Eq.~(\ref{rad7}) without the two factors
$\Theta(x_{n-1}^\prime-x_n)\Theta(x_n-x_n^\prime)$. Importantly, this
product is symmetric in $x_{n-1}^\prime$ and $x_n^\prime$. The second
derivatives yield
\begin{eqnarray}
\lefteqn{\frac{\partial}{\partial x_n}
\prod_{i> j}\Theta(x_i-x_j^\prime)\prod_{j\ge l}\Theta(x_j^\prime-x_l)=
      \prod(\Theta_{\neq nn^\prime} , \Theta_{\neq (n-1)^\prime n})}\cr
    &&\Big(\delta^\prime(x_n-x_n^\prime)\Theta(x_{n-1}^\prime-x_n)+
      \delta^\prime(x_{n-1}^\prime-x_n)\Theta(x_n-x_n^\prime)+\Big.\cr
    &&\Big.\delta(x_{n-1}^\prime-x_n)\delta(x_n-x_n^\prime)\Big)\quad .
\label{rad8}
\end{eqnarray}
The last term vanishes upon integration, since it is symmetric in
$x_{n-1}^\prime$ and $x_n^\prime$, whereas the rest of the integrand
is antisymmetric due to the Vandermonde determinant
$\Delta_{N-1}(x^\prime)$ in the measure (\ref{eq7.6}).
Differentiation with respect to $x_n^\prime$ gives
\begin{eqnarray}
\lefteqn{\frac{\partial}{\partial x_n^\prime}\prod_{i> j}\Theta(x_i-x_j^\prime)
         \prod_{j\ge l}\Theta(x_j^\prime-x_l)=}\cr
    &&\qquad\qquad\prod(\Theta_{\neq n^\prime(n+1)}, \Theta_{\neq nn^\prime})
      \Big(\delta(x_n^\prime-x_{n+1})\Theta(x_n-x_n^\prime) \Big.\cr
    &&\qquad\qquad\qquad\qquad\qquad
      \Big. -\delta(x_n-x_{n}^\prime)\Theta(x_n^\prime-x_{n+1})\Big) \ .
\label{rad9}
\end{eqnarray}
Integration by parts of the first term of the right hand side of 
Eq.~(\ref{rad6}) yields
\begin{eqnarray}
\lefteqn{\Delta_x\int \widetilde{\mu}(x^\prime,x) 
           \Phi_{N-1}^{(\beta)}(x^\prime,\widetilde{k})
           \prod_{i> j}\Theta(x_i-x_j^\prime)
           \prod_{j\ge l}\Theta(x_j^\prime-x_l)d[x^\prime]=}\cr 
 &&        \qquad\qquad\qquad\int \widetilde{\mu}(x^\prime,x)
           \Delta_{x^\prime} 
           \Phi_{N-1}^{(\beta)}(x^\prime,\widetilde{k})d[x^\prime]-  
           k_N^2\int \widetilde{\mu}(x^\prime,x)
           \Phi_{N-1}^{(\beta)}(x^\prime,\widetilde{k})d[x^\prime]\cr
 &&        +2 \int \Phi_{N-1}^{(\beta)}(x^\prime,\widetilde{k})
           \sum_{n=1}^{N-1}\Bigg(
           \prod(\Theta_{\neq n^\prime(n+1)} ,\Theta_{\neq nn^\prime})
            \Bigg.\cr
 &&        \qquad\qquad\Big(\delta(x_n-x_n^\prime)\Theta(x_n^\prime-x_{n+1})
           +\delta(x_n^\prime-x_{n+1})\Theta(x_n-x_n^\prime)\Big)\cr
 &&        \left.\left(\frac{\partial}{\partial x_n}+
           \frac{\partial}{\partial x_n^\prime}+
           \frac{1}{2}\sum_{m\neq n}\frac{\beta}{x_n-x_m}+
           -\frac{1}{2}\sum_{m\neq n}
           \frac{\beta}{x_n^\prime-x_m^\prime}\right)\right)
           \widetilde{\mu}(x^\prime,x)
           d[x^\prime]\quad .
\label{rad10}
\end{eqnarray} 
Inserting in Eq.~(\ref{rad10}) the function
$\widetilde{\mu}(x^\prime,x)$ as given in Eq.~(\ref{Arad3}) and
Eq.~(\ref{eq7.6}) we find after a straightforward calculation
\begin{eqnarray}
\lefteqn{\Delta_x\int \widetilde{\mu}(x^\prime,x) 
          \Phi_{N-1}^{(\beta)}(x^\prime,\widetilde{k})
          d[x^\prime]=}\cr 
 &&       \int \widetilde{\mu}(x^\prime,x)
          \Delta_{x^\prime} 
          \Phi_{N-1}^{(\beta)}(x^\prime,\widetilde{k})d[x^\prime]\,-\,
          k_N^2\int \widetilde{\mu}(x^\prime,x)
          \Phi_{N-1}^{(\beta)}(x^\prime,\widetilde{k})d[x^\prime]\cr
 &&       +2 \int \Phi_{N-1}^{(\beta)}(x^\prime,\widetilde{k})
          \sum_{n=1}^{N-1}\Bigg(
          \prod(\Theta_{\neq n^\prime(n+1)} , \Theta_{\neq nn^\prime})
          \bigg(g(x_n^\prime;y_n;x,x^\prime)\bigg. \Bigg. \cr
 &&       \qquad\qquad\qquad 
          \bigg.-g(x_n^\prime;x_n;x,x^\prime)\bigg) \cr
 &&       \Bigg.\Big(\delta(x_n-x_n^\prime)\Theta(x_n^\prime-x_{n+1})
          +\delta(x_n^\prime-x_{n+1})\Theta(x_n-x_n^\prime)\Big)\Bigg)
          \widetilde{\mu}(x^\prime,x) d[x^\prime]\quad ,
\label{rad11}
\end{eqnarray}
with
\begin{eqnarray}
g(x_n;x_n^\prime;x,x^\prime)&=&\left(\beta/2-1\right)
           \left(\sum_{m=1}^{N-1}\frac{1}{x_n-x_m^\prime}-
           \sum_{m\neq n}\frac{1}{x_n-x_m}\right) \cr
g(x_n^\prime;x_n;x,x^\prime)&=&\left(\beta/2-1\right)
           \left(\sum_{m\neq n}\frac{1}{x_n^\prime-x_m^\prime}-
           \sum_{m=1}^N\frac{1}{x_n^\prime-x_m}\right) \quad .
\label{rad12}
\end{eqnarray}
We now can perform the integration of the $\delta$ distributions in
Eq.~(\ref{rad11}). We notice that the difference
$(g(x_n^\prime;x_n;x,x^\prime)- g(x_n^\prime;x_n;x,x^\prime))$
vanishes linearly, whenever $x_n^\prime$ approaches one of the
boundaries of its integration domain. Thus the second integral in
Eq.~(\ref{rad11}) yields zero as long as the measure diverges less
than $(x_n-x_n^\prime)^{-1}$ when $x_n^\prime$ approaches $x_n$. This
is always the case for $\beta>0$.  Collecting everything, we arrive at
the identity~(\ref{eq7.8}).

\section{Symmetry of the Radial Functions 
                     for Arbitrary $\beta$}
\label{appSYR}
\renewcommand{\theequation}{E.\arabic{equation}}
\setcounter{equation}{0}

Applying the recursion formula~(\ref{eq7.5}) to all $N-1$ levels, we
can extend Eq.~(\ref{eq2.3a}) to arbitrary $\beta$ and write
\begin{eqnarray}
\Phi_N^{(\beta)}(x,k) &=&
       \int \prod_{n=1}^{N-1} d\mu(x^{(n)},x^{(n-1)}) 
                             \nonumber\\
     & & \qquad\qquad
       \exp\left(i\left(\sum_{m=1}^{N-n+1} x_m^{(n-1)}-
                     \sum_{m=1}^{N-n} x_m^{(n)}\right)k_{N-n+1}\right)
                             \nonumber\\
     & & \qquad\qquad \exp\left(ix_1^{(N-1)}k_1\right) 
\label{eqSYR.1}
\end{eqnarray}
where $x^{(0)}=x$.  Analogously, we also find
\begin{eqnarray}
\Phi_N^{(\beta)}(k,x) &=&
       \int \prod_{n=1}^{N-1} d\mu(k^{(n)},k^{(n-1)}) 
                             \nonumber\\
     & & \qquad\qquad
       \exp\left(i\left(\sum_{m=1}^{N-n+1} k_m^{(n-1)}-
                     \sum_{m=1}^{N-n} k_m^{(n)}\right)x_{N-n+1}\right)
                             \nonumber\\
     & & \qquad\qquad \exp\left(ik_1^{(N-1)}x_1\right) 
\label{eqSYR.2}
\end{eqnarray}
with $k^{(0)}=k$ for the solution of the differential equation which
results from Eq.~(\ref{eq7.1}) by interchanging $x$ and $k$. We have
to show that these two radial functions~(\ref{eqSYR.1})
and~(\ref{eqSYR.2}) are identical. To this end, we change in
Eq.~(\ref{eqSYR.1}) on the $n^{\rm th}$ level the variables
$x_m^{(n)}, \ m=1,\ldots,(N-n)$ to $k_m^{(n)}, \ m=1,\ldots,(N-n)$ by
setting
\begin{equation}
\frac{\prod_{l=1}^{N-n-1}(x_m^{(n-1)}-x_l^{(n)})}
            {\prod_{l\ne m}(x_m^{(n-1)}-x_l^{(n-1)})}
\ = \ r_m^{(n)} \ = \
\frac{\prod_{l=1}^{N-n-1}(k_m^{(n-1)}-k_l^{(n)})}
            {\prod_{l\ne m}(k_m^{(n-1)}-k_l^{(n-1)})} 
\label{eqSYR.3}
\end{equation}
for $n=1,\ldots,(N-1)$.  These are, on the $n^{\rm th}$ level, $N-n+1$
equations for making a change of $N-n$ variables. However, one has
\begin{equation}
\sum_{m=1}^{N-n+1} r_m^{(n)} \ = \ 1
\label{eqSYR.4}
\end{equation}
on all levels which eliminates one of the $N-n+1$ equations.

Of course, the substitution~(\ref{eqSYR.3}) is motivated by the radial
Gelfand--Tzetlin coordinates which we introduced to construct the
recursion formula for $\beta=1,2,4$. In this case, the $r_m^{(n)}$ are
the moduli squared of a column of a matrix $U \in U(N-n;\beta)$.
Here, we do not use this connection to matrices and groups. We simply
view Eq.~(\ref{eqSYR.3}) as a standard change of variables in an
integral. We underline that Eq.~(\ref{eqSYR.3}) does not involve
$\beta$ at all.  Equation~(\ref{eqSYR.4}) is just the normalization of
a column of $U$ for $\beta=1,2,4$. Since it is independent of $\beta$,
it also holds for arbitrary $\beta$. One can also verify
Eq.~(\ref{eqSYR.4}) by a direct calculation.

The original domains of integration are $x_m^{(n-1)} \le x_m^{(n)} \le
x_{m+1}^{(n-1)}$. In these boundaries, the $r_m^{(n)}$ are positive
definite. Hence, to satisfy this when changing the variables, we must
have $k_m^{(n-1)} \le k_m^{(n)} \le k_{m+1}^{(n-1)}$ for the new
domains of integration.

To work out the measure in the new variables $k_m^{(n)}$, we interpret
Eq.~(\ref{eqSYR.3}) as a change to the integration variables
$r_m^{(n)}$, too. This yields immediately
\begin{equation}
\frac{\Delta_{N-n}(x^{(n)})}{\Delta_{N-n+1}(x^{(n-1)})}
 \, d[x^{(n)}]
\ = \ d\mu(r^{(n)}) \ = \
\frac{\Delta_{N-n}(k^{(n)})}{\Delta_{N-n+1}(k^{(n-1)})} 
 \, d[k^{(n)}] \ .
\label{eqSYR.5}
\end{equation}
The first equality sign goes back to the radial Gelfand--Tzetlin
coordinates. We may use this piece of information, because it is
independent of $\beta$.  The second equality sign is simply due to
Eq.~(\ref{eqSYR.3}). Using this result, we find for the full and
$\beta$ dependent measure
\begin{eqnarray}
d\mu(x^{(n)},x^{(n-1)}) &=& G_{N-n+1}^{(\beta)} \, 
 \left( \frac{\prod_{l,m}(x_m^{(n-1)}-x_l^{(n)})}
             {\Delta_{N-n+1}^2(x^{(n-1)})} \right)^{(\beta-2)/2}
                           \nonumber\\
 & & \qquad\qquad\qquad\qquad
 \frac{\Delta_{N-n}(x^{(n)})}{\Delta_{N-n+1}(x^{(n-1)})}
 \, d[x^{(n)}]
                             \nonumber\\
 &=& G_{N-n+1}^{(\beta)} \, 
    \left( \prod_{m=1}^{N-n+1} r_m^{(n)} \right)^{(\beta-2)/2}
 \, \frac{\Delta_{N-n}(x^{(n)})}{\Delta_{N-n+1}(x^{(n-1)})}
 \, d[x^{(n)}]
                             \nonumber\\
 &=& G_{N-n+1}^{(\beta)} \, 
     \left( \frac{\prod_{l,m}(k_m^{(n-1)}-k_l^{(n)})}
             {\Delta_{N-n+1}^2(k^{(n-1)})} \right)^{(\beta-2)/2}
                           \nonumber\\
 & & \qquad\qquad\qquad\qquad
 \frac{\Delta_{N-n}(k^{(n)})}{\Delta_{N-n+1}(k^{(n-1)})}
 \, d[k^{(n)}]
                             \nonumber\\
 &=& d\mu(k^{(n)},k^{(n-1)}) \ . 
\label{eqSYR.6}
\end{eqnarray}
For $\beta=1,2,4$, this result is a direct consequence of the
invariance of the group measure $d\mu(U)$. Here, we have derived it
for arbitrary $\beta$. This, in turn, implies that the invariance of
the group measure $d\mu(U)$ is embedded into and reflects much more
general features.

We now collect all these intermediate results and plug them into
Eq.~(\ref{eqSYR.1}). Apart from the expressions in the exponential
functions, we have full agreement with the right hand side of
Eq.~(\ref{eqSYR.2}). Hence, it remains to be shown that the
change of variables~(\ref{eqSYR.3}) leads to the identity
\begin{eqnarray}
& & \sum_{n=1}^{N-1} \left(\sum_{m=1}^{N-n+1} x_m^{(n-1)}-
                           \sum_{m=1}^{N-n} x_m^{(n)}\right)k_{N-n+1}
         \, + \, x_1^{(N-1)}k_1
                             \nonumber\\
& & \qquad\qquad = \
    \sum_{n=1}^{N-1} \left(\sum_{m=1}^{N-n+1} k_m^{(n-1)}-
                           \sum_{m=1}^{N-n} k_m^{(n)}\right)x_{N-n+1}
         \, + \, k_1^{(N-1)}x_1 \ .
\label{eqSYR.7}
\end{eqnarray}
Since the symmetry relation~(\ref{eq7.3}) holds for $\beta=1,2,4$, we
know that Eq.~(\ref{eqSYR.7}) must be true in these cases. However, as
Eq.~(\ref{eqSYR.7}) does not involve $\beta$ at all, it must also be
valid for arbitrary $\beta$. Inserting this into the right hand side
of Eq.~(\ref{eqSYR.1}), we recover Eq.~(\ref{eqSYR.2}), as desired. We
notice that this line of arguing cannot be spoiled by any other
contribution to the argument of the exponential functions, because all
other terms in the integrand are purely algebraic. This completes the
proof of the symmetry relation~(\ref{eq7.3}) for arbitrary $\beta$.

\section{Calculation of the Normalization 
                     Constant $G_N^{(\beta)}$}
\label{appNKB}
\renewcommand{\theequation}{F.\arabic{equation}}
\setcounter{equation}{0}

In the previous App.~\ref{appSYR}, we introduced the coordinates
$r_n^\prime=r_n^{(1)}, \ n=1,\ldots,N$ on the first level of the
recursion. They are the moduli squared of the coordinates on the unit
sphere in the complex $N$ dimensional space. Thus, it is natural to
use the following type of hyper spherical coordinates
\begin{eqnarray}
\sqrt{r_n^\prime} &=& \cos\vartheta_n \, 
                      \prod_{\nu=1}^{n-1} \sin\vartheta_\nu \ ,
                      \qquad n=1,\ldots,(N-1) \ ,
                                     \nonumber\\
\sqrt{r_N^\prime} &=& \sin\vartheta_{N-1} \, 
                      \prod_{\nu=1}^{N-2} \sin\vartheta_\nu
\label{eqNKB.1}
\end{eqnarray}
where the positive semidefiniteness of the $r_n^\prime$ restricts the
domain of integration to $0 \le \vartheta_n < \pi/2, \ 
n=1,\ldots,(N-1)$. Thus, we integrate over a $(2^N)^{\rm th}$ segment
of the unit sphere. The measure
\begin{equation}
d\mu(r^\prime) \ = \ \prod_{n=1}^{N-1} \sin^{2(N-n)-1}\vartheta_n \,
                                       \cos\vartheta_n \, d\vartheta_n 
\label{eqNKB.2}
\end{equation}
is, apart from the phase angles, the measure on the unit sphere. 
Collecting everything, we have 
\begin{eqnarray}
1 &=& \int d\mu(x^\prime,x) \ = \
      G_N^{(\beta)} \, \int 
      \left(\prod_{n=1}^N \sqrt{r_n^\prime}\right)^{\beta-2} \,
      d\mu(r^\prime)                             \nonumber\\
  &=& G_N^{(\beta)} \, \prod_{n=1}^{N-1} 
      \int_0^{\pi/2} \sin^{(N-n)\beta-1}\vartheta_n \,
                     \cos^{\beta-1}\vartheta_n \,
                     d\vartheta_n                \nonumber\\
  &=& G_N^{(\beta)} \, \prod_{n=1}^{N-1} 
            \frac{\Gamma((N-n)\beta/2)\Gamma(\beta/2)}
                 {2\Gamma((N-n+1)\beta/2)} \ = \
      G_N^{(\beta)} \, \frac{\Gamma^N(\beta/2)}
                                  {2^{N-1}\Gamma(N\beta/2)}
\label{eqNKB.3}
\end{eqnarray}
where the integral over $\vartheta_n$ is just Euler's integral of the
first kind.

\section{Translation Invariance of 
                     $W_{N,\omega}^{(\beta)}(x,k)$}
\label{appWPD}
\renewcommand{\theequation}{G.\arabic{equation}}
\setcounter{equation}{0}

We shift every $x_n$ in the the recursion formula~(\ref{eq7.5}) for
arbitrary $\beta$ by a constant $\bar{x}$ and obtain
\begin{eqnarray}
\Phi_N^{(\beta)}(x+\bar{x},k) &=&
         \int d\mu(x^\prime,x+\bar{x}) \, 
  \exp\left(i\left(\sum_{n=1}^Nx + N\bar{x} - 
                  \sum_{n=1}^{N-1}x^\prime\right)k_N\right) \, 
                             \nonumber\\
    & & \qquad\qquad\qquad\qquad
         \Phi_{N-1}^{(\beta)}(x^\prime,\widetilde{k})
\label{eqWPD.1}
\end{eqnarray}
with $x_n+\bar{x} \le x_n^\prime \le x_{n+1}+\bar{x}$ as the domains
of integration. The change of variables $x_n^\prime\longrightarrow
x_n^\prime+\bar{x}$ removes $\bar{x}$ from the measure given in
Eq.~(\ref{eq7.6}) and the domains of integration, we find
\begin{eqnarray}
\Phi_N^{(\beta)}(x+\bar{x},k) &=& \exp\left(i\bar{x}k_N\right) \,
         \int d\mu(x^\prime,x) \, 
  \exp\left(i\left(\sum_{n=1}^Nx - 
                  \sum_{n=1}^{N-1}x^\prime\right)k_N\right) 
                             \nonumber\\
    & & \qquad\qquad\qquad\qquad
         \Phi_{N-1}^{(\beta)}(x^\prime+\bar{x},\widetilde{k}) \ .
\label{eqWPD.2}
\end{eqnarray}
We want to employ an induction. We assume that the radial functions
for arbitrary $\beta$ have the property
\begin{equation}
\Phi_N^{(\beta)}(x+\bar{x},k) \ = \ 
      \exp\left(i\bar{x}\sum_{n=1}^Nk_n\right) \,
      \Phi_N^{(\beta)}(x,k) \ .
\label{eqWPD.3}
\end{equation}
If this is correct for $N-1$, formula~(\ref{eqWPD.2}) implies that it
is also true for $N$. The induction starts with $N=2$ where the
correctness of Eq.~(\ref{eqWPD.3}) is immediately obvious from the
explicit solution~(\ref{eq1.31}) for arbitrary $\beta$. Thus, 
Eq.~(\ref{eqWPD.3}) is valid for all $N$.

Since the $k_n$ are arbitrary and since the sum over all $k_n$ is
invariant under the permutations $\omega(k)$, the
property~(\ref{eqWPD.3}) must also be true for every function
$\Phi_{N,\omega}^{(\beta)}(x,k)$ with $\omega\in S_N$.  We compare
this with the expression
\begin{equation}
\Phi_{N,\omega}^{(\beta)}(x+\bar{x},k) \ = \ 
      \exp\left(i\bar{x}\sum_{n=1}^Nk_n\right) \,
       \frac{\exp\left(i\sum_{n=1}^Nx_nk_{\omega(n)}\right)}
            {|\Delta_N(x)\Delta_N(k)|^{\beta/2}} \,
                             W_{N,\omega}^{(\beta)}(x+\bar{x},k) 
\label{eqWPD.4}
\end{equation}
which results from the Hankel ansatz~(\ref{eq1.32}). Hence, we
conclude that we necessarily have
\begin{equation}
W_{N,\omega}^{(\beta)}(x+\bar{x},k) 
 \ = \ W_{N,\omega}^{(\beta)}(x,k) \ .
\label{eqWPD.5}
\end{equation}
This is the translation invariance.

\section{Calculation of $\Phi_4^{(4)}(x,k)$}
\label{appG}
\renewcommand{\theequation}{H.\arabic{equation}}
\setcounter{equation}{0}

We perform the calculation for $\Phi_4^{(4)}(-ix,k)$ to avoid
inconvenient factors of $i$. The operator $L_{x,\omega(k)}$ defined in
Eq.~(\ref{eq1.34}) splits into two parts.  The first part
\begin{equation}
\widetilde{\Delta}_{x,\omega(k)} \ = \ \sum_{n=1}^N
                                    \frac{\partial^2}{\partial x_n^2}-
                                    4\sum_{n<m}\frac{1}{(x_n-x_m)^2}\quad .
\label{G1}
\end{equation}
does not change the order in $k$, while the second one,
\begin{equation}
\Lambda_{x,\omega(k)} \ = \ 2 \sum_{n=1}^N
       k_{\omega(n)}\frac{\partial}{\partial x_n} \ , 
\label{G2}
\end{equation}
raises the order in $k$ by one.  Since we can restrict ourselves to
one element of the permutation group, we discuss only the identity
permutation in the sequel.  The symmetry of $x$ and $k$ together with
the result for $\Phi_3^{(4)}(x,k)$ suggests one to try an expansion in
the composite variable $z_{ij}$ as defined in Eq.~(\ref{usp9}).  To
this end we define the elementary symmetric functions
\begin{equation}
e_{\nu}(z) \ = \ \sum_{i_1j_1 < i_2j_2<\ldots<i_\nu j_\nu}
                 \prod_{l=1}^\nu z_{i_lj_l} \ .
\label{G3}
\end{equation}
Here, we assume the following ordering of the composite index
$\left\{i_lj_l\right\}$, $i_l<j_l$. We say $\left\{i_lj_l\right\} <
\left\{i_mj_m\right\}$ if $i_l < i_m$ or $i_l = i_m$ and $j_l < j_m$.
  All indices run to $N$.  The highest order elementary symmetric
  function is of order $N(N-1)/2$ and is given by
  $\Delta_N(x)\Delta_N(k)$. The asymptotic formula (\ref{eq1.34aa})
  yields the leading term for large arguments. It is the starting
  point for a recursion in powers of $z^{-1}$.
\begin{equation}
W_N^{(4)}(z) \ = \ \sum_{\nu=0}^{N(N-1)/2} p_\nu(z^{-1})\quad ,
\label{G4}
\end{equation}
where $p_\nu(z)$ is a symmetric function  of order
$\nu$ in $x_i$ and $k_i$. We investigate the action of
the two operators defined in Eq.~(\ref{G1}),(\ref{G2}) and find
\begin{eqnarray}
\Lambda_{x,k} e_{\nu}(z^{-1}) & = & -2 \sum_{n<m}^N\frac{1}{(x_n-x_m)^2} 
                            e_{\nu-1}(z_{\neq nm}^{-1})
                            \label{G5}\\
\widetilde{\Delta}_{x,k} e_{\nu}(z^{-1}) & = & 
                             -4\; \sum_{n<m}^N\frac{1}{(x_n-x_m)^2}
                             e_{\nu}(z_{\neq nm}^{-1})\cr
                       &   & -2\; \sum_{n<m \atop {k\neq n \atop k\neq m}}^N
                             \frac{1}{(x_n-x_m)^2}\;
                             z_{nk}^{-1}\;
                             z_{mk}^{-1}\;
                             e_{\nu-2}(z_{\neq nm 
                             \atop {\neq nk
                             \atop \neq mk}}^{-1})\quad .
\label{G6}
\end{eqnarray}
The function $e_{\nu}(z_{\neq nm})$ is the elementary symmetric
function $e_{\nu}(z)$ with all terms containing $z_{nm}$ omitted.  For
$\nu =0,1,2$ we simply have $p_\nu(z^{-1})=(-2)^\nu e_\nu(z^{-1})$.
For $\nu \geq 3$ the last term in Eq.~(\ref{G5}) causes corrections to
the elementary symmetric functions. This arises due to the mixed
derivatives which have to be taken into account in the action of
$\widetilde{\Delta}_{x,k}$ onto $e_\nu(z^{-1})$ for $\nu\geq 3$.
Because of this term the Hankel Ansatz seems becomes increasingly
cumbersome as higher values of $N$ are considered, since more and more
correction terms have to be constructed. So far, the construction was
only possible for $N = 4$.  To construct the correction terms
explicitly for the case $N=4$, we define a new set of symmetric
functions as follows
\begin{equation}
f_{\nu}(z^{-1}) \ = \ \sum_{k<l<m}^N z_{kl}^{-1}
                              z_{km}^{-1}
                              z_{lm}^{-1}
                             e_{\nu-3}(z^{-1}_{\neq kl 
                             \atop {\neq km
                             \atop \neq lm}})\quad .
\label{G7}
\end{equation}
Again we have to investigate the action of $\Lambda_{x,k}$ and 
$\widetilde{\Delta}_{x,k}$ on  $f_{\nu}(z^{-1})$. We find
\begin{equation}
\widetilde{\Delta}_{x,k}\,f_{3}(z^{-1}) \ = \ 
                             -4 \sum_{n<m}^N\frac{1}{(x_n-x_m)^2}
                             \;f_3(z_{\neq nm}^{-1})
\label{G7a}
\end{equation}
and
\begin{equation}
 \Lambda_{x,k}\, f_3(z^{-1}) \ = \
                             -2 \sum_{n<m \atop {k\neq n \atop k\neq m}}^N
                             \frac{1}{(x_n-x_m)^2}\;
                             z_{nk}^{-1}\;
                             z_{mk}^{-1} \ ,
\label{G8}
\end{equation}
thus $f_3(z^{-1})$ is the desired correction term. We have
\begin{equation}
p_3(z^{-1}) \ = \ -2^3\left(e_3(z^{-1})+\frac{1}{2}\,
                   f_3(z^{-1})\right)\quad .
\label{G9}
\end{equation}
Fortunately, due to Eq.~(\ref{G7a}) in the next step the correction
term itself has not to be corrected and we find
\begin{equation}
p_4(z^{-1}) \ = \ 2^4\left(e_4(z^{-1})+\frac{1}{2}\, 
                  f_4(z^{-1})\right) \ .
\label{G10}
\end{equation}
Up to now these results are valid for arbitrary $N$.  The action of
$\widetilde{\Delta}_{x,k}$ onto the symmetric function $f_4(z^{-1})$
 is not as simple as Eq.~(\ref{G7a}). After a series of manipulations
 we arrive at
\begin{eqnarray}
\widetilde{\Delta}_{x,k}\,f_{4}(z^{-1}) &=&
                         -4\;\sum_{n<m}^N\frac{1}{(x_n-x_m)^2}
                         \;f_4(z_{\neq nm}^{-1}) \\
                         & & \qquad\qquad 
                         - 2 \sum_{n<m \atop {k\neq n \atop k\neq m}}^N
                         \frac{1}{(x_n-x_m)^2}\;
                         z_{nk}^{-1}\;
                         z_{mk}^{-1}\;
                         f_{2}(z_{\neq nm 
                         \atop {\neq nk
                         \atop \neq mk}}^{-1}) \ .
\label{G11}
\end{eqnarray}
The contribution (\ref{G6}) has to be added to this expression
stemming from the action of $\widetilde{\Delta}_{x,k}$ onto 
$e_4(z^{-1})$. On the other hand we calculate 
\begin{eqnarray}
\Lambda_{x,k}\, f_5(z^{-1}) &=&
                      -2\;\sum_{n<m}^N\frac{1}{(x_n-x_m)^2}
                      \;f_4(z_{\neq nm}^{-1}) \\
                      & & \qquad\qquad 
                      - 2 \sum_{n<m \atop {k\neq n \atop k\neq m}}^N
                      \frac{1}{(x_n-x_m)^2}\;
                      z_{nk}^{-1}\;
                      z_{mk}^{-1}\;
                      e_{2}(z_{\neq nm 
                      \atop {\neq nk
                      \atop \neq mk}}^{-1}) \ .
\label{G12}
\end{eqnarray}
Thus, we have to find yet another correction term to compensate the
second term in Eq.~(\ref{G11}). We define
\begin{equation}
f_5^\prime(z^{-1}) \ = \ \sum_{i_1<i_2<i_3<i_4}\;\prod_{r<j}
                      z_{i_ri_j}^{-1}\;
                      \sum_{r<j}z_{i_ri_j}
                     =\ \sum_{j<k \atop l<m} 
                      z_{jl}^{-1}z_{jm}^{-1}
                      z_{kl}^{-1}z_{km}^{-1}
                      z_{lm}^{-1}
\label{G13}
\end{equation}
and see that $\Lambda_{x,k}\, f_5^\prime(z^{-1})$ yields exactly the
desired second term of Eq.~(\ref{G11}).  Pushing forward this
procedure becomes more complicated step by step.  There seems to be no
obvious way of constructing the additional terms. Apparently for
higher orders the correction terms also involve an increasing amount
of indices.  Nevertheless for $N=4$ we are already at the end of the
recursion.  Then the general expression
\begin{equation}
p_5(z^{-1})=- 2^5\left(e_5(z^{-1})+\frac{1}{2}\, f_5(z^{-1})
                 +\frac{1}{4}\,f_5^\prime(z^{-1})\right)
\label{G14}
\end{equation}
reduces to 
\begin{equation}
p_5(z^{-1})= - 72\,e_5(z^{-1})\quad .
\label{G15}
\end{equation}
The last step can readily be done, since the action of  
$\widetilde{\Delta}_{x,k}$ onto $e_5(z^{-1})$ is already known by 
Eq.~(\ref{G6}). Thus we arrive at
\begin{equation}
p_6(z^{-1})= 288\, e_6(z^{-1})\quad .
\label{G16}
\end{equation}
Importantly, we have 
\begin{equation}
\widetilde{\Delta}_{x,k}\,e_6(z^{-1}) = 
\widetilde{\Delta}_{x,k}\,\frac{1}{\Delta_4(x)\Delta_4(k)} = 0\quad .
\label{G17}
\end{equation}
That means, the sequence finishes after the sixth step.  Collecting
everything and observing that, for $N=4$, $f_5(z)= 2e_5(z)$ and
$f_6(z)= 4 e_6(z)$, we obtain
\begin{equation}
W_4^{(4)}(x,k)= \sum_{\nu=1}^6 (-2)^\nu e_\nu(z^{-1}) 
                + \sum_{\nu=3}^6 (-2)^{\nu-1} f_\nu(z^{-1})
                - 8\;e_5(z^{-1}) + 96\;e_6(z^{-1}) \ .
\label{G18}
\end{equation} 
This can be rewritten more in a more compact way as
\begin{eqnarray}
W_4^{(4)}(x,k)&=& \frac{1}{\Delta_4(x)\Delta_4(k)}\Bigg(
              \prod_{i<j}(2- z_{ij}) + \Big.\cr
           &  &
          \Bigg.\frac{1}{2}\sum_{l<m<n}
             \prod_{i<j \atop {\neq lm \atop {\neq ln \atop \neq mn}}}
              (2- z_{ij}) + \frac{1}{4}
              \sum_{l<m \atop k<n}
\prod_{i<j\ \neq lk \ \neq ln \atop {\neq mk \ \neq mn
          \ \neq kn}}(2-z_{ij})\Bigg) 
\label{G19}
\end{eqnarray}
which yields Eq.~(\ref{usp11}).


\begin{thebibliography}{30}

\bibitem{AS}          Abramowitz, M., Stegun, I.A.: 
                      Handbook of Mathematical Functions,
                      9th edition.
                      New York: Dover Publications, 1970
\bibitem{BEE}         Beenakker, C.W., Rejaei, B.: Exact Solution for 
                      the Distribution of Transmission Eigenvalues
                      \ldots.
                      Phys. Rev. B {\bf 49}, 7499 (1994)
\bibitem{CAL}         Calogero, F: Solution of a Three--Body Problem
                      in One Dimension.
                      J. Math. Phys.{\bf 10} 2191 (1969)
\bibitem{CAS}         Caselle, M.: Distribution of Transmission
                      Eigenvalues in Disordered Wires.
                      Phys. Rev. Lett. {\bf 74}, 2776 (1995)
\bibitem{DH}          Duistermaat, J.J.:, Heckman, G.J.: On the
                      Variation in the Cohomology of the Symplectic
                      Form of the Reduced Phase space.
                      Inv. Math. {\bf 69}, 259 (1982)
\bibitem{DYS1}        Dyson, F.J.: Statistical Theory of the Energy 
                      Levels of Complex Systems. I,II,III.
                      J. Math. Phys. {\bf 1}, 140 (1962)
\bibitem{DYS2}        Dyson, F.J.: A Brownian-Motion Model for the
                      Eigenvalues of a Random Matrix.
                      J. Math. Phys. {\bf 1}, 1191 (1962)
\bibitem{FOR}         Forrester, P.J.: Integration Formulas and exact
                      Calculations in the Calogero--Sutherland Model.
                      Mod. Phys. Lett. B. {\bf 9}, 359 (1995)
\bibitem{GEL1}        Gelfand, I.M.: Spherical functions on 
                      symmetric Riemann spaces.
                      Dokl. Akad. Nauk. SSSR {\bf 70} 5 (1950)
\bibitem{GT}          Gelfand, I.M., Tzetlin, M.L.: Matrix Elements
                      for the Unitary Groups (Russian). 
                      Dokl. Akad. Nauk. {\bf 71}, 825 (1950)
\bibitem{Gil}         Gilmore, R.:
                      Lie Groups, Lie Algebras and some of
                      their Applications.
                      New York: John Wiley \& Sons, 1974
\bibitem{GK1}         Gross, K.I., Kunze, R.A.: Bessel Functions and
                      Representation Theory I. 
                      J. Funct. Anal. {\bf 22}, 73 (1976)
\bibitem{GK2}         Gross, K.I., Kunze, R.A.: Bessel Functions and
                      Representation Theory II.
                      J. Funct. Anal. {\bf 25}, 1 (1977)
\bibitem{GGT}         Guhr, T.: Gelfand--Tzetlin Coordinates 
                      for the Unitary Supergroup.
                      Commun. Math. Phys. {\bf 176}, 555 (1996) 
\bibitem{GuKo}        Guhr, T., Kohler, H.: Recursive Construction 
                      for a Class of Radial Functions I --- 
                      Superspace.
                      Submitted for Publication in Communications
                      of Mathematical Physics.
\bibitem{GMGW}        Guhr, T., M\"uller--Groeling, A.,
                      Weidenm\"uller, H.A.: Random Matrix Theories in 
                      Quantum Physics: Common Concepts.
                      Phys. Rep. {\bf 299}, 189 (1998)
\bibitem{HA}          Ha, Z.N.C.: Fractional Statistics in one
                      Dimension \ldots .
                      Nuc. Phys. B. {\bf 435}, 604 (1995)
\bibitem{Haake}       Haake, F.: Quantum Signatures of Chaos.
                      Berlin: Springer Verlag, 1991
\bibitem{HC1}         Harish-Chandra: Differential Operators on a 
                      Semisimple Lie Algebra.
                      Am. J. Math. {\bf 79}, 87 (1957)
\bibitem{HC}          Harish-Chandra: Spherical Functions on a
                      Semisimple Lie Group I. 
                      Am. J. Math. {\bf 80}, 241 (1958)
\bibitem{HEL}         Helgason, S.: Groups and Geometric Analysis.
                      San Diego: Academic Press, 1984
\bibitem{HER}         Hertz, C.S.: Bessel Functions of Matrix
                      Arguments. 
                      Ann. of Math. {\bf 61} 474 (1955)
\bibitem{HOL}         Holman, W.J.: Generalized Bessel functions 
                      and the representation theory of 
                      $U(2)\sigma C^{2\times 2}$.
                      J. Math. Phys. {\bf 21}, 1977 (1980)
\bibitem{Hua}         Hua, L.K.: Harmonic Analysis of Functions of 
                      Several Complex Variables in the Classical
                      Domains.
                      Providence: American Physical Society, 1963.
\bibitem{IZ}          Itzykson, C., Zuber, J.B.: The Planar
                      Approximation II.
                      J. Math. Phys. {\bf 21}, 411 (1980) 
\bibitem{KON}         Kontsevich, M.: Intersection Theory on the
                      Moduli Space of Curves and the Matrix
                      Airy Functions.
                      Commun. Math. Phys. {\bf 147}, 1 (1992)
\bibitem{Mehta}       Mehta, M.L.: Random Matrices, 2nd edition.
                      San Diego: Academic Press, 1990
\bibitem{MUI}         Muirhead, R.J.: Aspects of multivariate
                      statistical Theory.
                      New York: Wiley, 1982
\bibitem{NF}          Nagao, T., Forrester, P.J.: Correlations for the
                      circular Dyson Brownian motion model \ldots.
                      Nuc. Phys. B. {\bf 532}, 733 (1998)
\bibitem{OO}          Okounkov, A., Olshanski, G.: Shifted Jack
                      Polynomials, Binomial Formula, and 
                      Applications.
                      Math. Res. Lett. 4, 69 (1997).
\bibitem{OP}          Olshanetsky, M.A., Perelomov, A.M.: 
                      Quantum Integrable Systems Related to Lie
                      Algebras. 
                      Phys. Rep. {\bf 94}, 313 (1983)
\bibitem{GO}          Olshanski, G.: private communication.
\bibitem{PAN}         Pandey, A.: Brownian--motion Model of Discrete
                      Spectra.
                      Chaos, Sol. and Frac. {\bf 5}, 1275 (1995)
\bibitem{RIS}         Risken, H: The Fokker--Planck Equation.
                      Berlin: Springer--Verlag, 1989
\bibitem{SLS}         Shatashvili, S.L.: Correlation Functions
                      in the Itzykson-Zuber Model. 
                      Commun. Math. Phys. {\bf 154}, 421 (1993)
\bibitem{STA}         Stanley, R.P.: Some Combinatorial Properties of
                      Jack Symmetric Functions.
                      Adv. Math. {\bf 77}, 76 (1989)
\bibitem{SU1}         Sutherland, B.: 
                      Exact results for a Quantum Many--Body Problem in 
                      One Dimension II. 
                      Phys. Rev. A. {\bf 5}, 1372 (1972) 
\bibitem{SZA}         Szabo, R.J.: Equivariant Localization of Path
                      Integrals.
                      preprint:{\sc hep-th/9608068}
\bibitem{UL1}         Ullah, N.: J. Math. Phys. {\bf 4}, 1279 (1963).
\bibitem{UL2}         Ullah, N.: Matrix Ensembles in the
                      Many--Nucleon Problem.
                      Oxford: Clarendon Press, 1987

\end{thebibliography}
\end{document}